\documentclass[aps,pra,reprint,groupedaddress,superscriptaddress,10pt,longbibliography]{revtex4-2}
\pdfoutput=1 
\usepackage[utf8]{inputenc}
\usepackage{graphicx}
\usepackage{orcidlink}
\usepackage{booktabs}
\usepackage{tabularx}
\usepackage{cancel}

\usepackage{siunitx} 
\newcommand{\omatt}[1]{}

\usepackage{soul} 
\usepackage[normalem]{ulem}
\usepackage{amssymb}
\usepackage{physics}
\usepackage{hyperref}
\DeclareSIUnit\bohr{\text{\ensuremath{a_\textup{0}}}}

\begin{document}

\title{Exposing and overcoming the polarizability pathology of the Fermi-Omont pseudopotential}

\author{S. Efe G\"urleyen
\orcidlink{0000-0003-3363-3202}}
\email{Contact author: sgurleye@purdue.edu}
\affiliation{Department of Physics and Astronomy, Purdue University, West Lafayette, IN 47907, USA}

\author{Matthew T. Eiles
\orcidlink{0000-0002-0569-7551}}
\affiliation{Department of Physics and Astronomy, Purdue University, West Lafayette, IN 47907, USA}

\author{Chris H. Greene
\orcidlink{0000-0002-2096-6385}}
\affiliation{Department of Physics and Astronomy, Purdue University, West Lafayette, IN 47907, USA}
\affiliation{Purdue Quantum Science and Engineering Institute, Purdue University, West Lafayette, IN 47907, USA}

\date{\today}

\begin{abstract}
Any realistic electron-atom or ion-atom potential energy function contains a long-range contribution from the atomic polarizability which gives a $1/r^4$ long-range tail. This modifies the well-known Wigner threshold law for orbital momenta $L>0$, causing these $L\neq 0$ scattering phase shifts to vary linearly in energy instead of as $k^{2L+1}$. The $L>0$ Fermi-Omont zero-range pseudopotentials are erroneously divergent at zero energy due to this modification, a divergence deriving from very long-range physics that should not be incorporated into a Dirac delta function. Here, we investigate some alternative scenario for repairing this pathological behavior to obtain smooth and accurate potential energy curves calculated using the Fermi-Omont pseudopotential concept, with some systematic benchmarking tests for comparison.
\end{abstract}

\maketitle
\section{Introduction}

The power of the Fermi-Omont zero-range pseudopotential\cite{Fermi1934,omont1977theory} to simplify the theoretical description of quantum mechanical phenomena in diverse fields of physics has had some noteworthy successes~\cite{FermiNeutronScattering,BECtheory,osti_4322571,ChrisTrilobite}. Fermi's original treatment showed that, in the long-wavelength limit, the exact interaction between two particles, in the absence of Coulomb forces, can be replaced by a Dirac delta function pseudopotential whose coefficient is proportional to the zero energy $S$-wave scattering length between those particles. In atomic units (a.u.), Fermi's pseudopotential is given by
\begin{align}\label{spseudo}
    V_0(\vec{r};\vec{R})=2\pi a_0(\varepsilon)\delta^{(3)}(\vec{r}-\vec{R}).
\end{align}
where $\varepsilon$ is the kinetic energy of the scattering electron, $a_0(\varepsilon)$ is the energy dependent scattering length and $\vec{R}$ is the position of the scatterer atom. 

Omont's treatment generalized this to higher partial waves, and the $P$-wave term in particular,
\begin{align}\label{Ppseudo}
    V_1(\vec{r};R)=6\pi a_1^3(\varepsilon)\overleftarrow{\nabla}\cdot\delta^{(3)}(\vec{r}-\vec{R})\overrightarrow{\nabla},
\end{align}
has found extensive use~\cite{hamilton2002shape,ObservationPend,Venu2023}. Here, $a_1^3(\varepsilon)$ is the energy dependent scattering volume and $\overrightarrow{\nabla}(\overleftarrow{\nabla})$ are the gradient operators acting on the right(left) hand side.

Some technical difficulties involving these singular potentials were later circumvented by minor modifications that recast them in a regularized form~\cite{busch_two_1998,PhysRevA.61.063416,christrap,PhysRevA.71.050701,PhysRevA.79.062701,highpartialwaves,PhysRevA.81.052710}, but with the same basic idea of using them to enforce the correct long-range scattering phase shifts in the lowest partial waves that are derived from the true or exact potential functions, or in some cases, directly from experiment~\cite{MilenaMatt}.

Despite the successes of many applications of the Fermi-Omont pseudopotentials, it has become increasingly apparent that some failures arise in quantitative applications. It is the purpose of the present study to address those failures head on, to analyze them, and to propose a strategy to alleviate many of the fundamental difficulties that arise in practice. 

In short, although a zero-range pseudopotential can be suitable for replacing short-range potentials in the low energy limit, the present study documents how it is insufficient to represent long- or even intermediate-range potentials by itself. Our strategy to overcome this fundamental issue derives from this observation. Firstly, the long-range potential  is truncated at some finite distance in our treatment by a cutoff function. In this way, the conflict between the pseudopotential and the long-range potential is alleviated. Secondly, the physics beyond the cutoff distance has to be incorporated, and we show that our proposed resolution can address that long-range physics in most cases through a perturbative correction. 

\section{Preliminary considerations about phase shift behavior in long-range Potentials}

One of the clearest indications that there are fundamental difficulties with the Fermi-Omont theory arises when a long-range polarization potential $-\frac{\alpha}{2 r^4}$ is present.  Specifically, consider the behavior of the $P$-wave $(L = 1)$ pseudopotential in the zero energy limit.  Recall first that for a short- or finite-range potential, the Wigner threshold law implies that every partial-wave $L$ phase shift varies at low energy as $\delta_L(k) \rightarrow  -C_L k^{2L+1}$, where in a.u., $k = (2 \varepsilon)^{1/2}$ and, for the $P$-wave, the constant $C_L$ is related to the scattering volume, $C_1 \equiv a_1^3(k) \equiv -\frac{\tan{\delta_1(k)}}{k^3}$ as $k\to 0$. For short-range potentials, therefore, the scattering volume is finite as $\varepsilon\to 0$. However, we know from~\cite{omalleypol,LowEngScatLRPot} that the low energy phase shift in a polarization potential is quadratic in $k$ for all $L\ge 1$,
\begin{equation}\label{leading term}
    \delta_L(\varepsilon) = \frac{\pi \alpha}{(2L-1)(2L+1)(2L+3)} k^2,\,\,\varepsilon\to 0.
\end{equation}
This implies that the the zero-energy scattering volumes for all $L\ge 1$ diverge in proportion to $1/k^{2L-1}$.
When an electron interacts with a heavy neutral atom, as an example, taking Rb that we study in this work, the $P$-wave phase shift is determined by two separate contributions.  There is a short-range interaction that incorporates the electron-atom physics of screening and exchange with the core electrons and is limited in range to just a few Bohr radii. Additionally, there is the long-range polarization interaction that extends to $r\rightarrow \infty$; the combined centrifugal plus long-range potential has a local maximum at a distance of $r_{\rm max}=\sqrt{\alpha}$. For the Rb ground state, $r_{\rm max} = 17.9\,a_0$ and $V(r_{\rm max})= 42.7$ meV, corresponding to $k=0.056\,a_0^{-1}$.  

Another key distance to keep in mind is the outermost classical turning point, which is close to $r_{\rm outer} \sim (L+\frac{1}{2})/k$ for energies below the barrier maximum, that diverges at zero-energy. This is relevant because for that below-barrier energy range, the phase shift is primarily determined at distances $r>r_{\rm outer}$ (unless there is an electronic shape resonance).  These different distance ranges are mentioned here also because they provide some insight into why the Fermi-Omont pseudopotential fails to encapsulate the energy content of the electron-scatterer interaction in this energy range.

Yet another energy range which poses a different challenge theoretically is the negative electron-atom collision energies, {\it i.e.} when $\varepsilon=(ik)^2 \equiv -\kappa^2 <0$ with $k,\kappa$ both positive by convention.  In order to smoothly transition from positive energy solutions into solutions in this energy range it is useful to define a scattering phase shift relative to an energy-analytic base pair of solutions. Because the negative energy scattering phase shifts, {\it i.e.} referred to the analytic reference functions, are somewhat unfamiliar, we review some of the key aspects in their behavior and contrast them with the more standard energy-normalized solutions. The regular and irregular energy-normalized solution pair $(f_{\varepsilon L}(r),g_{\varepsilon L}(r))$ are, at positive energy $\varepsilon >0$,
\begin{align}
    f_{\varepsilon L}(r)&\equiv  \sqrt{\frac{2}{\pi k}} kr j_L(kr), \\
    g_{\varepsilon L}(r)&\equiv  \sqrt{\frac{2}{\pi k}}kr y_L(kr),
\end{align}
where $j_L(k r), y_L(k r)$ are the regular and irregular spherical Bessel functions, respectively.
The energy-normalized phase shifts are defined in terms of the Wronskians ($\rm Wr$) of the radial wavefunction solution $F_{\varepsilon L}(r)$ with this base pair, {\it i.e.} as
\begin{equation}\label{engNormPS}
    \delta_{\varepsilon L}= \lim_{r\to \infty}\arctan{\frac{{\rm Wr}(f_{\varepsilon L},F_{\varepsilon L})}{{\rm Wr}(g_{\varepsilon L},F_{\varepsilon L})}}.
\end{equation}
For a finite-range potential, the solutions are matched at a finite radius $r_M$ that exceeds the range of the scattering potential. If it instead involves an infinite range component, then the matching radius extends to infinity for zero-energy scattering. In practice, as long as the potential decays faster than $1/r$, converged phase shifts can be achieved by doing the matching at a very large but finite distance.

If the problem takes significant contributions from the negative kinetic energy domain of the electron-scatterer collision, energy-normalized phase shifts cannot be used, as they are not well-defined in this energy range (except as complex quantities through analytic continuation).
 On the other hand, the energy-analytic base pair of solutions, $(f_{\varepsilon L}^{\rm an}(r),g_{\varepsilon L}^{\rm an}(r))$, are defined as in~\cite{GQDTI,GQDTII} by
\begin{align}
    f_{\varepsilon L}^{\rm an}(r)&\equiv (k r_0)^{-L-1/2} \sqrt{\frac{2}{\pi k}} kr j_L(kr), \\
    g_{\varepsilon L}^{\rm an}(r)&\equiv (k r_0)^{L+1/2} \sqrt{\frac{2}{\pi k}}kr y_L(kr),
\end{align}
where $r_0$ can be chosen as any convenient length scale. 
These solutions to the radial free-particle Schr\"odinger equation remain smooth and analytic as the energy decreases through 0, with the continuation $k \rightarrow i \kappa$. As a result, the energy-analytic scattering phase shifts are the smooth (over the full energy domain) counterparts of the more familiar phase shifts relative to energy-normalized regular and irregular solutions, and are analytic functions of energy only for a finite range potential that decays asymptotically faster than any inverse power law~\cite{GQDTI}. The phaseshift information in the Fermi-Omont pseudopotential is thus better characterized by the analytic phase shifts.

For a radial potential that is of short-range, the analytic phase shift is obtained with respect to the analytic base pair
\begin{equation}\label{engAnPS}
    \delta^{\rm an}_{\varepsilon L}= \lim_{r\to\infty} \arctan{\frac{{\rm Wr}(f^{\rm an}_{\varepsilon L},F_{\varepsilon L})}{{\rm Wr}(g^{ \rm an}_{\varepsilon L},F_{\varepsilon L})}}.
\end{equation}
Again, in computational applications, the solutions are matched outside the short-range potential's range, at a radius $r_M$ as done for energy-normalized phase shifts. The branch of the arctangent function is chosen to maintain continuity, which also ensures smoothness at all energies. The energy-analytic and energy-normalized phase shifts are related to each other for positive energies by $\tan\delta_{\varepsilon L}^{\rm an}=\tan\delta_{\varepsilon L}/(k r_0)^{2L+1}=(a_L(\varepsilon)/r_0)^{2L+1}$.

Some aspects of the behavior of negative energy scattering phase shifts are discussed in~\cite{GQDTI,PhysRevA.20.656}, and are analyzed here.  First of all, the electron-scatterer Schr\"odinger equation in partial wave $L$ does not possess a negative energy solution unless the phase shift obeys the following bound state quantization condition,
\begin{equation}\label{bound state curve}
\tan\delta_{\varepsilon L}^{\rm an}+(-1)^L( \kappa r_0)^{-2L-1}=0.
\end{equation}
Although this condition is rarely satisfied in near zero-energy scatterings, it still can be utilized to estimate the phase shift or scattering length for slightly negative energies.

In the spirit of quantum defect theory, and closely related effective range theory, as well as R-matrix theory, one often deals with exponentially divergent negative energy solutions that are analytically continued from positive to negative energies. A scattering phase shift at slightly negative energy can still make sense and be computed, {\it e.g.} using Eq.~\eqref{engNormPS}, for a short-range potential function. But for any long-range power law potential such as the polarizability potential relevant to electron-neutral scattering, the phase shift at $\varepsilon<0$ is not even well-defined mathematically. The pathological behavior can be seen by inspecting the integral equation for the radial wavefunction, namely:
\begin{equation}
    F_{\varepsilon L}(r) = f_{\varepsilon L}^{\rm an}(r) + \pi \int_0^\infty dr' f_{\varepsilon L}^{\rm an}(r_<)g_{\varepsilon L}^{\rm an}(r_>)V(r') F(r').
\end{equation}
It is clear that since both $f_{\varepsilon L}^{\rm an}(r)$ and $g_{\varepsilon L}^{\rm an}(r)$ diverge exponentially at infinity, so does $F_{\varepsilon L}(r)$ at any non-bound energy.  Consequently, the integral needed to define the analytic phase shift, namely 
\begin{equation}
    \tan{\delta_{\varepsilon L}^{\rm an}} = -\pi \int_0^\infty dr' f_{\varepsilon L}^{\rm an}(r') V(r') F_{\varepsilon L}(r') 
\end{equation}
is exponentially divergent.

One final remark about zero-range pseudopotentials, in general, is their failure to represent mid- or long-range interactions. It tries to encapsulate the physical information of the exact potential that spreads out to large distances to a single point on the space, {\it i.e.} the position of the scatterer. We demonstrate the consequence of this remark in the potential energy curves of an ultra long-range Rydberg molecule as a model study and present a method to partially repair them.

\section{Our Strategy and an Application}\label{chap:stratandapp}

Our strategy begins by regularizing the unphysical divergence of the scattering volume at zero-energy. This is done by eliminating the source of this issue, that is, the long-range tail of the polarization potential, by introducing a long-range(LR) cutoff $f_{\rm LR}(r;r_{\rm LR})$ to the potential, $\bar{V}_{\rm Sc}(r) \equiv V_{\rm Sc}(r)f_{\rm LR}(r;r_{\rm LR})$. The cutoff function is chosen as $f_{\rm LR}(r;r_{\rm LR})=\frac{1}{2}[1-\tanh(\frac{r-r_{\rm LR}}{\Delta r})]\in[0,1]$ where $r_{\rm LR}\in[50,150] \, a_0$ is the cutoff radius and $\Delta r=5 \,a_0$.

One caveat is that the long-range physics in the outer region beyond the cutoff radius is often non-negligible, and contributions coming from the peeled off long-range part, {\it i.e.} $\Delta V_{\rm Sc}=V_{\rm Sc}(r)[1-f_{\rm LR}(r;r_{\rm LR})]$ should be taken into account with a plausible treatment. As we will show, there is typically -- depending on the scattering energy -- a range of cutoff radii such that the effect of the long-range potentials can be included perturbatively.  

The outmost well of a Rydberg Rb$(np)$-Rb$(5s)$ dimer in its molecular $\Sigma$ state typically lies in the region around the zero electron collision energy divergence. Therefore, it is our choice for applying our ideas as they naturally furnish a system where our regularization strategy is crucial to obtain smooth potential energy curves that are otherwise divergent and cuspy.
\subsection{A model system: Rubidium dimer}

Consider a 3-body problem consisting of a positively charged ionic core at the origin, a Rb($np$) Rydberg electron at $\vec{\rho}$, and a ground state perturbing Rb$(5s)$ atom at $\vec{R}=R\hat{z}$ with which the electron collides in $\Sigma$ state, forming the states $np\Sigma$. Let $\vec{r}=\vec{\rho}-\vec{R}$ be the electron's relative coordinate to the scatterer(perturber).

The $e^-$-Rb$(n p)$ potential is chosen locally from the $l=0$ part of the nonlocal potential in Eq.~(18) of Ref.~\cite{PhysRevA.49.982}:
\begin{align}\label{electronCorePot}
    V_C(\rho)=-\frac{Z_C(\rho)}{\rho}-\frac{\alpha}{2\rho^4}(1-e^{-(\rho/\rho_C)^6}),
\end{align}
where $Z_C(\rho)=1+(\mathcal{Z}-1)e^{-a_1 \rho}-\rho(a_3+a_4\rho)e^{-a_2\rho}$. The values of the parameters are the same as in the reference and are given in Table~\ref{tab:parameterqd}. The energy levels calculated with this local potential have an error rate of $\lesssim 0.2\%$ compared to the energy levels of $np$ and $ns$ states in the NIST database~\cite{NIST_RbI_2024}. The $j$ dependent energy levels of the same $p$ states are averaged in this error analysis.

\begin{table}[h!]
\centering

\begin{tabular*}{0.6\columnwidth}{@{}l@{\extracolsep{\fill}}r@{}}
\toprule
\multicolumn{2}{c}{%
  \parbox{0.5\columnwidth}{\centering\bfseries
  Model potential parameters (from Ref.~\cite{PhysRevA.49.982})}} \\
\midrule
$a_1$         & 3.69628474  \\
$a_2$         & 1.64915255  \\
$a_3$         & -9.86069196 \\
$a_4$         & 0.19579987  \\
$\rho_C$      & 1.66242117  \\
$\mathcal{Z}$ & 37          \\
\bottomrule
\end{tabular*}

\vspace{3mm}

\begin{tabular*}{0.6\columnwidth}{@{}l@{\extracolsep{\fill}}r@{}}
\toprule
\multicolumn{2}{c}{\textbf{Quantum defect parameters}} \\
\midrule
$\mu_0(0)$  & 3.13159  \\
$\mu'_0(0)$ & 0.17331  \\
$\mu_1(0)$  & 2.66235  \\
$\mu'_1(0)$ & 0.28255  \\
$\mu_2(0)$  & 1.57115  \\
$\mu'_2(0)$ & -0.08178 \\
$\mu_3(0)$  & 0.020163 \\
$\mu'_3(0)$ & -0.12295 \\
\bottomrule
\end{tabular*}
\caption{Model potential and quantum defect parameters.}
\label{tab:parameterqd}

\end{table}

The quantum defects are extracted from the energy levels of this potential and they are parameterized as
\begin{align}
    \mu_l(n)=\mu_l(0)+\frac{\mu'_l(0)}{(n-\mu_l(0))^2}.
\end{align}
The values of these parameters are given in Table~\ref{tab:parameterqd}. The electron's angular momentum relative to the core, $l$, is not to be confused with the electron's angular momentum relative to the perturber, $L$. 

The local $e^-$-Rb($5s$) potential is adopted from the $^3 S_1$ component of the nonlocal potential of Ref.~\cite{khuskivadze2002adiabatic} with the addition of our long-range cutoff function:
\begin{align}\label{ScatPot}
    \bar{V}_{\rm Sc}(r;r_{\rm LR})=\bigg[&-\frac{\alpha}{2r^4}(1-e^{-(r/r_{\rm SR})^6})&\nonumber\\&-\frac{A}{r}e^{-\gamma r}\bigg]f_{\rm LR}(r;r_{\rm LR}).
\end{align}
Here, $r_{\rm SR}$ is a short-range cutoff for the polarization potential. The parameters of the scattering potential are chosen as $ A=66.08, \gamma=9.33, r_{SR}=1.8834, \alpha=319.2 $ because it yields similar $^3S_1$- and $^3P_1$-wave scattering phase shifts with Ref.~\cite{khuskivadze2002adiabatic} with this parameter set although our potential is a local adaptation of the one in the reference. 

When the electron in its Rydberg state governed by these core and perturber potentials collides with the perturber, it has a very low kinetic energy, $\varepsilon =\frac{k^2}{2}= -\frac{1}{2\nu^2}+\frac{1}{R}$, and a long de Broglie wavelength during the collision ({\it e.g.} in this work, $k\sim 0.0095\,a_0^{-1}$ a.u. for $30p\Sigma$ and $k\sim  0.017\,a_0^{-1}$  for $20p\Sigma$). Hence, the scattering is primarily influenced by the $S$-wave, and occasionally by the $P$-wave because the low energy collisions usually cannot penetrate the centrifugal barriers beyond the $P$-wave to interact with the perturber's core through higher partial waves(remember that the outer maximum of the effective $e^-$-Rb$(5s)$ potential is reached at $k=0.056\,a_0^{-1}$). Furthermore, the long-wavelength feature of the collision also makes it suitable for the employment of pseudopotentials. For the purposes of the present study, which is focused on the accuracy and errors in computing the molecular potential curves, the kinetic terms associated with nuclear motion are neglected, referred to as the \textit{Born-Oppenheimer approximation}: the internuclear distance $R$ is treated as a parameter.  Moreover, all electronic and nuclear spins and the magnetic interactions are ignored, except that our calculations are based entirely on the triplet ($S=1$) total electronic spin of the Rb dimer.

As mentioned in the last paragraph, this system is a natural one for utilizing the zero-range Fermi-Omont pseudopotential that accounts for the $S$- and $P$-wave scatterings, given by (in a.u.):
\begin{align}
    \hat{V}_{\rm Sc}(\vec{\rho};R,r_{\rm LR})=2\pi\Big[&a_0(k;r_{\rm LR})\delta^{(3)}(\vec{\rho}-R\hat{z})\nonumber\\&+3 a_1^3(k;r_{\rm LR})\overleftarrow{\nabla}\cdot\delta^{(3)}(\vec{\rho}-R\hat{z})\overrightarrow{\nabla}\Big].
\end{align}
The scattering information obtained by means of the  potential in Eq.~\eqref{ScatPot} is stored in the scattering length and volume, namely $a_0(k)$ and $a_1^3(k)$, and computed using generalized quantum defect theory as summarized in Section~\ref{chap:EngAnPhSh}. The long-range cutoff dependence in the full potential is transferred to the scattering length/volume of the pseudopotential. One of the aims of this work is to investigate how the validity of modeling the exact potential with the Fermi-Omont pseudopotential changes as we vary the cutoff radius. It is expected to be more accurate for smaller electron-perturber scattering energies. Thereby, we study different Rb$(np)$-Rb$(5s)$ manifolds for which the outermost potential well has a smaller electron-perturber scattering energy for larger $n$.

Our long-range cutoff treatment kills the long-range tail of the polarizability interaction that lies between the electron-perturber barrier maximum at and zero energy; therefore, it is important to remember some key distances in diatomic ultralong-range Rydberg molecule(ULRM) systems in this regime. We already mentioned that the electron collision energy with the perturber at a distance $R$ from the Rydberg atom's nucleus is given semiclassically in a.u. as $\varepsilon = -\frac{1}{2 \nu^2}+\frac{1}{R}$. At $R=2\nu^2$, when the electron collision energy is zero, the pseudopotential modeling an infinite range polarizability potential diverges and leads to cusps in the Born-Oppenheimer potential energy curves(PECs). For an $L$-wave collision, the outer maximum of the scattering potential is at $r^L_{\rm max}=\sqrt{\frac{2\alpha}{L(L+1)}}$ which yields the value $V^L_{\rm sc}(r_{\rm max})=\frac{L^2(L+1)^2}{8\alpha}$. The resulting internuclear distance is then $R^L_{\rm min}=\left[\frac{[L(L+1)]^2}{8\alpha}+\frac{1}{2\nu^{2}}\right]^{-1}$. Below this distance, the electron's kinetic energy exceeds the barrier energy. So, in the interval $R\in\left[\frac{2\nu^{2}}{1+[L(L+1)\nu]^2/4\alpha},2\nu^{2}\right],$
the electron's kinetic energy is between the $L$-wave barrier and zero. The width of the corresponding length window is $\frac{2\nu^{2}}{1+4\alpha/[L(L+1)\nu]^2}$. This below-barrier collision regime is encountered for the $P$-wave in Rb($n p$)+Rb($5s$) collisions with effective principal quantum numbers in the range $\nu = 15, 30, 60$ at internuclear distances beyond (respectively): $R>264, 471, 586$ a.u. and it is the region that is mainly influenced by our cutoff treatment.

\subsection{Energy-analytic phase shifts under the cutoff treatment}\label{chap:EngAnPhSh}

We briefly mentioned the importance of the energy-analytic phase shifts for our strategy. In this section, we show how they behave under our long-range cutoff treatment, especially near- and below zero-energy. The energy-analytic phase shifts are computed using single-channel generalized quantum defect theory according to Eq.~\eqref{engAnPS}. We have chosen the arbitrary length factor $r_0$ to be the characteristic length scale of the polarization potential, {\it i.e.} $r_0=\sqrt{\alpha}\approx 17.9\, a_0$. 

\begin{figure}[h!]
\centering
\includegraphics[width=0.48\textwidth]{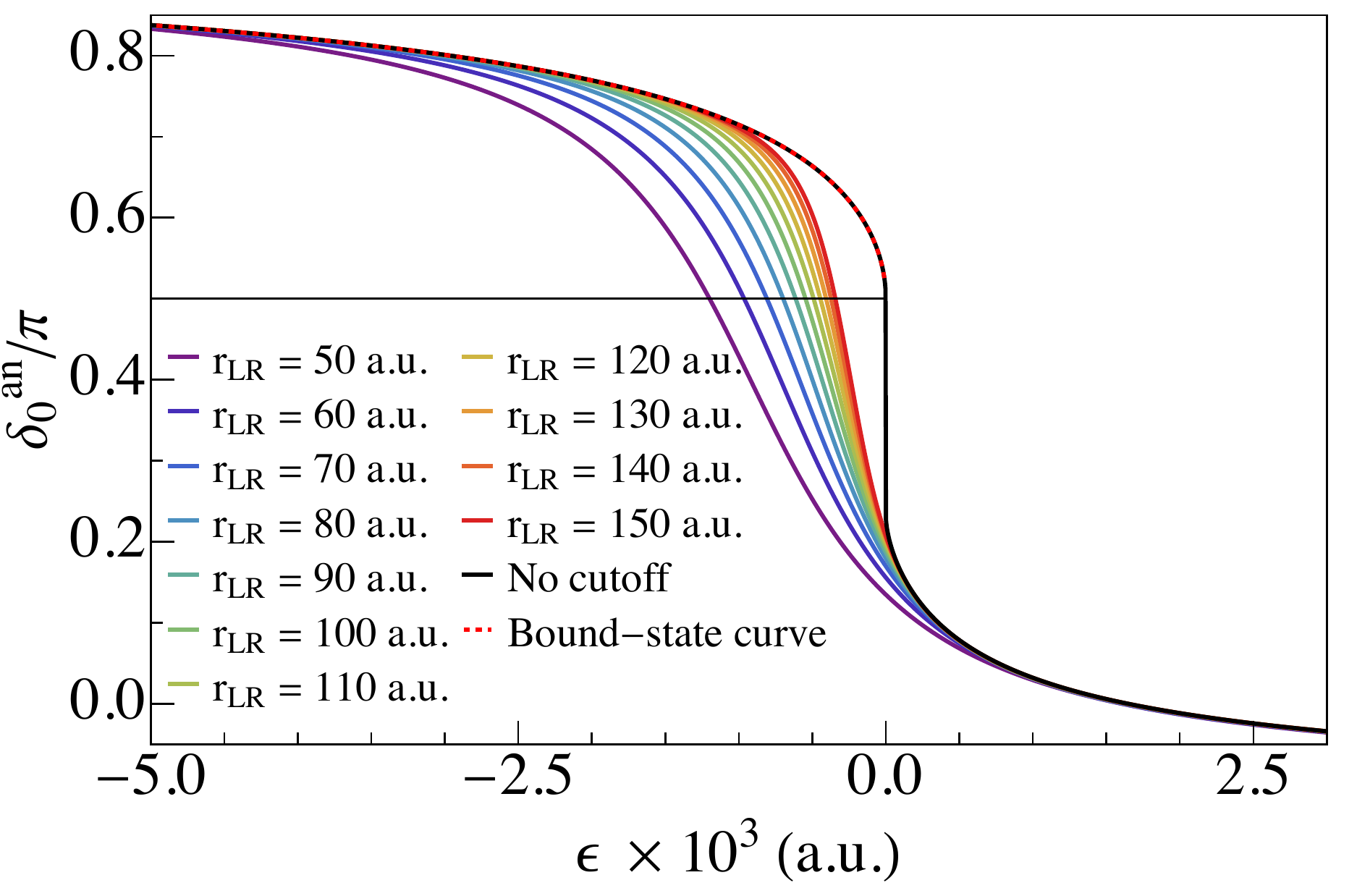}
\caption{The $S$-wave energy-analytic phase shifts of $e^-$-Rb($5s$) subjected to the potential in Eq.~\eqref{ScatPot} vs. the electron's collision energy; calculated with Eq.~\eqref{engAnPS} with different long-range cutoffs. The horizontal black curve marks the $\frac{\pi}{2}$ line which the scattering length diverges on; {\it i.e.} $\delta_0^{\rm an}(\varepsilon^*_{r_{\rm LR}};r_{\rm LR},r_0)=\frac{\pi}{2}$. This divergence is translated towards more negative energies as the cutoff radius gets shorter. The horizontal distance $\abs{\varepsilon^*_{r_{\rm LR}}}$ from the zero-energy where $\delta_0^{\rm an}(\varepsilon^*_{r_{\rm LR}};r_{\rm LR},r_0)=\frac{\pi}{2}$ occurs, obeys a power law: $\abs{\varepsilon^*_{r_{\rm LR}}}={0.1}r_{\rm LR}^{-1.131}$.}
\label{fig:Sphsh}
\end{figure}

\begin{figure}[h!]
\centering
\includegraphics[width=0.48\textwidth]{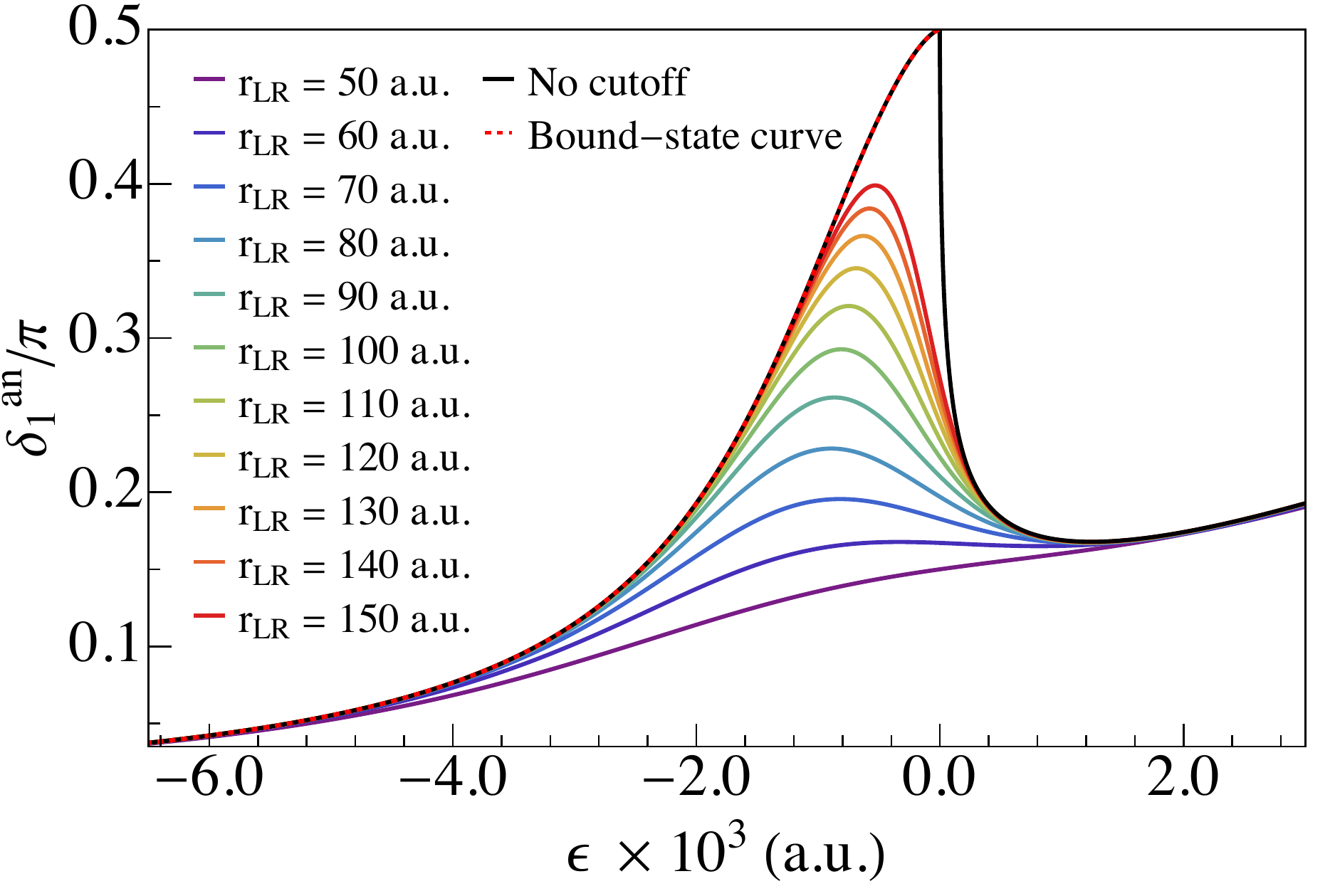}
\caption{The $P$-wave energy-analytic phase shifts of $e^-$-Rb($5s$) subjected to the potential in Eq.~\eqref{ScatPot} with different long-range cutoff radii vs. electron collision energy; calculated with Eq.~\eqref{engAnPS}. The long-range cutoff completely eliminates the zero-energy divergence as the energy-analytic phase shift never reaches the point $\delta_1^{\rm an}=\pi/2$ for any $r_{\rm LR}$.}
\label{fig:Pphsh}
\end{figure}

The $S$- and $P$-wave energy-analytic scattering phase shifts in an $e^-$-Rb($5s$) collision under the long-range cutoff treatment are presented in Figs.~\ref{fig:Sphsh} and \ref{fig:Pphsh}, respectively. We additionally show the $D$- and $F$-wave phase shifts in Fig.~\ref{fig:DFphsh} in Appendix \ref{append:DF}. These are calculated using the scattering potential given in Eq.~\eqref{ScatPot} with the same parameters in the following paragraph of that equation.

The red dashed curves in these phase shift plots show the bound state condition curve given in Eq.~\eqref{bound state curve}. This condition is rarely satisfied, since an atom like Rb has only one bound state in a deeper energy range than is considered for typical ULRMs, but as seen in Figs.~\ref{fig:Sphsh}, \ref{fig:Pphsh} and \ref{fig:DFphsh}, the phase shifts of every cutoff radii approach this curve and eventually follow it closely at some negative energy and below. Therefore, although we do not do it in this paper, this bound state curve can be utilized to estimate the negative energy phase shifts.

Note that the Fermi-Omont pseudopotential diverges when the analytic phase shift for any partial wave is equal to $\pi / 2$. A long-range cutoff pushes the $\delta_L^{\rm an}(\varepsilon^*_\infty=0;r_{\rm LR}=\infty,r_0)=\frac{\pi}{2}$ point of the even partial wave phase shifts from zero-energy to lower energies  $\delta_0^{\rm an}(\varepsilon^*_{r_{\rm LR}};r_{\rm LR},r_0)=\frac{\pi}{2}$(see Figs.~\ref{fig:Sphsh}, \ref{fig:DFphsh}) that follows a power law fitted to $\abs{\varepsilon^*_{r_{\rm LR}}}={0.1}r_{\rm LR}^{-1.131}$ for the $S$-wave. 

Furthermore, the cutoff function changes the Wigner threshold law obeyed by the potential back to the one for the finite-range potentials, $\delta_L\propto -C_L k^{2L+1} $. Consequently, the scattering volume goes to a constant instead of diverging at zero-energy, giving PECs that are free from this unphysical zero-energy divergence that occurs due to the infinite range of the polarizability interaction. It is worth noting that a very long cutoff radius can still lead to ill-behaving PECs because although $\tan\delta_1^{\rm an}$ does not reach the singular point, it still gets very large when $\delta_1$ approaches close values to $\pi/2$. The convergence towards the full-range potential results, which we denote here as the 'no cutoff' results and the satisfaction of $\delta^{\rm an}_L(\varepsilon;r_{\rm LR})\neq\pi/2$ for odd $L$ are apparent in Figs.~\ref{fig:Pphsh}, \ref{fig:DFphsh}.

As the long-range cutoff radius gets shorter, the inner region scattering potential becomes less attractive while as it gets longer, we can see that the phase shifts converge towards the most attractive no cutoff ({\it i.e.} infinite cutoff radius) case.
This effect is seen in the phase shifts presented, specifically as an overall decrease of the phase shifts as the cutoff radius decreases. Below 50 a.u. cutoff the behavior of the phase shift changes significantly because the cutoff radius gets too close to critical radii such as the $P$-wave barrier local maximum.

The same phase shifts are used throughout this paper, in particular when we apply the ideas in this study to an Rb$(n p)$+Rb$(5 s)$ collision. The usage of scattering phaseshifts at negative energies in a non-Coulomb potential is a key idea used in quantum defect theory; the negative energy behavior is discussed at length in Ref.\cite{PhysRevA.20.656}.
\subsection{Repairing the divergences in the PECs}

\begin{figure}[h!]
\centering
\includegraphics[width=0.48\textwidth]{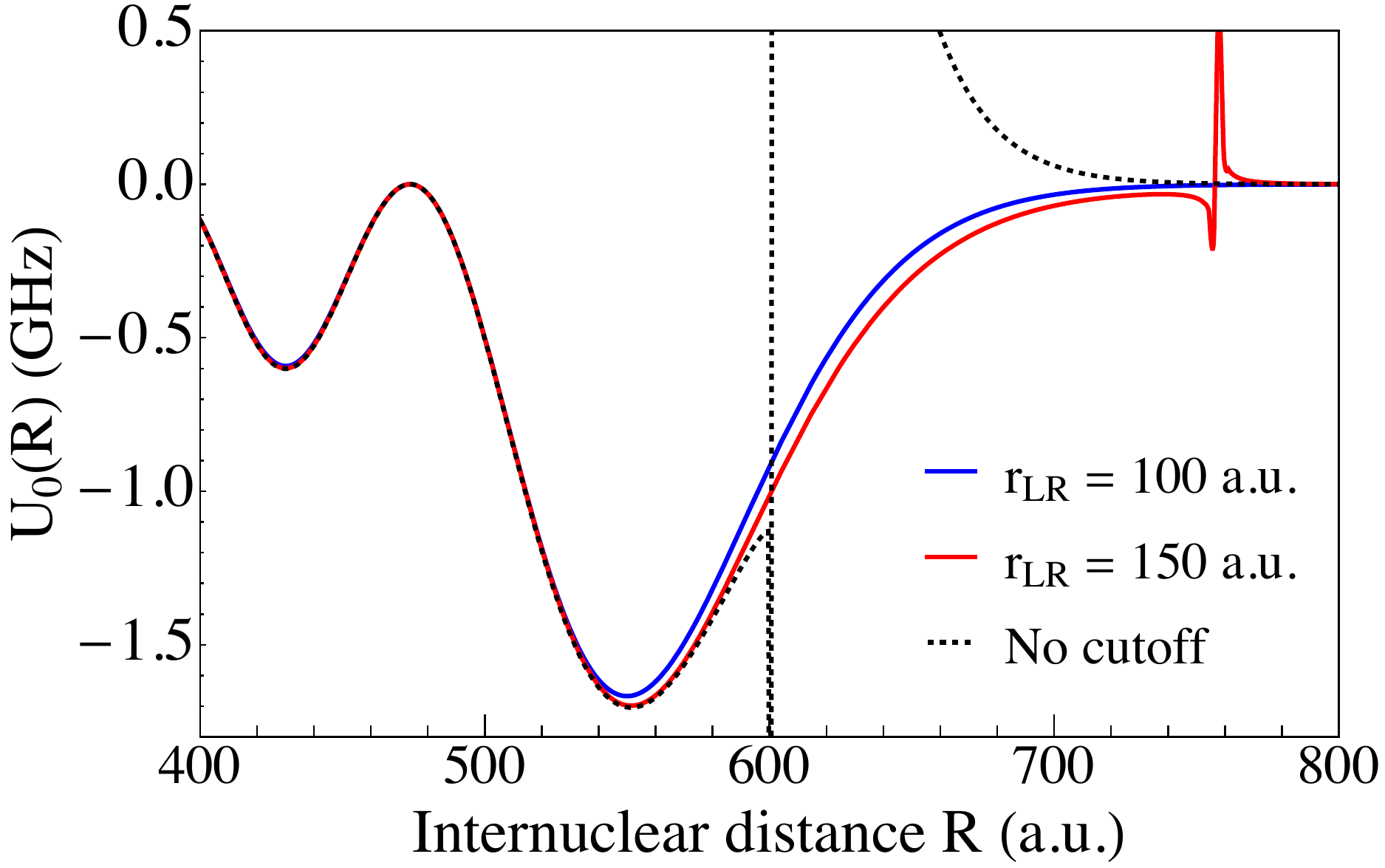}
\caption{Example PECs of an Rb$(np)$-Rb$(5s)$ dimer in its molecular $\Sigma$ state, calculated with the phase shifts that were presented in Fig.~\ref{fig:Sphsh} using Eq.~\eqref{SwavePECeq}. The no cutoff(full-range) curve displays the divergence due to $\delta_0^{\rm an}(0;\infty;\sqrt{\alpha})=\frac{\pi}{2}$ at $R_c(\nu,\infty)=2\nu^2$. Picking $r_{\rm LR}=150a_0$ pushes this divergence to $\delta_0^{\rm an}(\varepsilon^*_{150};150;\sqrt{\alpha})=\frac{\pi}{2}$ at $R_c=(\varepsilon^*_{150}+1/2\nu^2)^{-1}$ where $\varepsilon^*_{r_{\rm LR}}=-0.1r_{\rm LR}^{-1.131}$. This figure shows that it is possible to avert this divergence in the PECs that exist when the singular point $\delta_0^{\rm an}(\varepsilon^*_{r_{\rm LR}};r_{\rm LR},r_0)=\frac{\pi}{2}$ is near the outmost well by picking a shorter cutoff; in this case $r_{\rm LR}=100$ a.u.. This leads to a divergence at $R_c\approx 896a_0$, far away from the outmost well.}
\label{fig:Sdiv} 
\end{figure}

To demonstrate the significance of our long-range cutoff treatment to the PECs in the simplest way, let us employ only the $S$-wave component of the Fermi-Omont pseudopotential and invoke first order perturbation theory for the Rubidium dimer we introduced in this chapter. This approach leads to the potential energy curves given by
\begin{align}\label{SwavePECeq}
    U_0(R;r_{\rm LR})=\frac{3r_0}{2}\tan\delta_0^{\rm an}(\varepsilon;r_{\rm LR},r_0)\left(\frac{\mathcal{W}_{\nu,l+\frac{1}{2}}(R)}{R}\right)^2.
\end{align}
which are plotted in Fig.~\ref{fig:Sdiv} for the cases without a cutoff and with $r_{\rm LR}=100,150 a_0$ in the asymptotically $20p\Sigma$ state. The normalized Whittaker function in Eq.~\eqref{SwavePECeq} is 
\begin{equation}\label{NormWhittaker}
    \mathcal{W}_{\nu,l+\frac{1}{2}}(R)\equiv\frac{W_{\nu,l+\frac{1}{2}}\left( \frac{2R}{\nu}\right)}{\left[\nu^2 \Gamma(\nu+l+1)\Gamma(\nu-l)\right]^{1/2}},
\end{equation}
where $W_{\nu,l+\frac{1}{2}}\left(\frac{2R}{\nu}\right)$ is the standard Whittaker function. The quantum defects used here are given in Table~\ref{tab:parameterqd}. The same quantum defects are used throughout the rest of the paper, most importantly in the Green's function method presented in Section~\ref{chap:comments on on GFM}, and the perturbative correction calculations with eigenvectors that are represented by the Whittaker functions in Eq.~\eqref{NormWhittaker} with $l=1$.

Recall that the scattering length diverges when the phase shift takes the value $\pi/2$ at an energy $\varepsilon^*_{r_{\rm LR}}=-0.1r_{\rm LR}^{-1.131}$ as seen in Fig.~\ref{fig:Sphsh}. Therefore, in an $S$-wave scattering, when the electron sees the critical point $\delta_0^{\rm an}(\varepsilon^*_{r_{\rm LR}};r_{\rm LR},r_0)=\frac{\pi}{2}$ near the outmost well for some $r_{\rm LR}$, one can pick a shorter cutoff radius to which the critical point is pushed towards larger internuclear distances(smaller negative collision energies $\varepsilon$), far away from the well. The wavefunction does not vanish until $\delta_0^{\rm an}(\varepsilon^*_{r_{\rm LR}};r_{\rm LR},r_0)=\frac{\pi}{2}$, so the PEC diverges on this point. But the point $R_c(\nu,r_{\rm LR})$ which this divergence occurs depends on the cutoff radius. Therefore, if this divergence is located at a point that distorts the outmost well, it is possible to repair it by picking a shorter cutoff radius that makes $\delta_0^{\rm an}(\varepsilon^*_{r_{\rm LR}};r_{\rm LR},r_0)=\frac{\pi}{2}$ satisfied at an $R_c(\nu,r_{\rm LR})$ that is irrelevant to the outmost well, as shown in Fig.~\ref{fig:Sdiv}. $R_c$ can be estimated by plugging in the power law obtained into the semiclassical kinetic energy of the electron: $R_c=\left(-0.1r_{\rm LR}^{-1.131}+\frac{1}{2\nu^2}\right)^{-1}$.

In Fig.~\ref{fig:Sdiv}, of course the no cutoff potential as the most attractive one, yields the deepest well while the blue curve with the shortest cutoff radius is the shallowest. The difference between the curves starts to grow as the electron collision energy reaches  where the phase shifts of different cutoff radii in Fig.~\ref{fig:Sphsh} start to separate from each other, $\varepsilon\sim 0.00025$ a.u. The differences between the PECs without a cutoff and with different cutoff radii are addressed in the next chapter by taking into account the long-range cut off part of the potential through perturbation theory.

\section{Results: Rubidium Dimer}\label{chap:results}

Our ultimate aim is to verify the general validity of our PEC regularization strategy by eliminating the divergences in the energy-analytic phase shifts by means of the long-range cutoff function and check if the initial no cutoff PECs are obtained after adding back the long-range corrections perturbatively. Therefore, we compare a Green's function method(GFM) that uses the Fermi-Omont pseudopotential idea with a diagonalization treatment that uses the exact scattering potential; hence we refer it as full diagonalization(Diag.) in this paper. We show that the GFM that uses the phase shifts of the cut off potential plus the long-range corrections does not agree with the GFM itself that uses the full long-range potential phase shifts, but rather it agrees with the full diagonalization of the full potential.

When comparing the results of the two different methods, full diagonalization and GFM, one has to consider that the GFM takes into account only the $S$- and $P$-wave partial waves for the scattering of the Rydberg electron from the ground state Rb atom, whereas the full diagonalization considers all partial waves. Therefore, there are two main factors in determining the agreement between the two methods with the same potentials: (i) the level of accuracy of representing the exact potential by the pseudopotential; and (ii) the contributions of $L\geq 2$ partial waves coming from the inner region($r<r_{\rm LR}$). Both of these factors are decided by the length of the long-range cutoff radius. In this paper, we consider only the impacts of $D$- and $F$-waves out of all the $L\geq 2$ partial waves because the $L=3$ corrections are already very small(about $\lesssim 3$ MHz) in the outmost well for appropriately chosen cutoff radii as shown in Fig.~\ref{fig:PECcorrN202530}. They are estimated by first order perturbation theory using the higher order terms of Omont's pseudopotential expansion Eq.~(2.18) of Ref.~\cite{omont1977theory} as corrections to the GFM results by employing higher order pseudopotentials within perturbation theory. The details of this calculation are presented in Appendix~\ref{append:DF}. Their estimation is important in determining what portion of the disagreement between the two methods they possess and what portion comes from the Fermi-Omont pseudopotential of $S$- and $P$-waves.

The different regions in the problem require different resolutions as the electron's wave length varies due to the varying interaction strengths of the core and the perturber Rb atoms as the electron moves. Its wave function has rapid oscillations near the atomic cores and a long wave length when it is far from both of the cores. So, it is computationally convenient to have the ability to locally tune the basis set in different regions. The B-splines basis perfectly matches this condition; hence, it is our choice for the full diagonalization.

Now that we are ready to apply our general strategy, it is useful to summarize it here before we apply it in the next sections. Our general scheme of making the results of the different treatments that use a long-range exact potential and that rely on pseudopotentials modeling this long-range potential agree is as follows: (i) Since the long-range potential cannot be encapsulated to a point interaction, its range is cut by a smooth cutoff function. The consequent cutoff potential is suitable to be represented by a zero-range pseudopotential. (ii) Calculate molecular PECs using the pseudopotential-based treatment({\it e.g.} GFM) with the cut off potential to obtain smooth PECs(Rb$(np)$-Rb$(5s)$ near their outmost well in this work). (iiia) Take into account the peeled off long-range part of the exact potential with a suitable method; our choice in this paper was perturbation theory. (iiib) We additionally computed perturbatively the impact of $D$- and $F$-wave scatterings in the inner region, $r<r_{\rm LR}$, in this paper, because the GFM treatment ignores them and they are somewhat relevant in the PECs we investigate, especially the lower $n$ ones.

\subsection{Full diagonalization with B-splines and the accuracy of long-range corrections}

Here, we show that the cut off potential with the perturbative long-range correction gives the original potential's result using the B-spline basis setup given in Appendix~\ref{append:BspBasis}. The same setup is used for all the full diagonalization calculations in this paper. Let $U_{\rm B}(R),\bar{U}_{\rm B}(R;r_{\rm LR})$ be the PECs derived from the scattering potentials $V_{\rm Sc}(r)$ or $\bar{V}_{\rm Sc}(r;r_{\rm LR})$, respectively, using full diagonalization with a B-spline basis set and the quantity $\expval{\Delta V_{\rm Sc}}(R;r_{\rm LR})$ be the long-range perturbative correction to the Born-Oppenheimer potential curve caused by the long-range electron-perturber potential that had been removed when computing $\bar{U}_{\rm B}(R;r_{\rm LR})$. The perturbative corrections are calculated with Whittaker functions in Eq.~\eqref{NormWhittaker}, or by numerically calculated eigenvectors with full diagonalization for comparison, where we used the atomic effective principal quantum numbers $\nu$ for the Whittaker functions. Using perturbed($R$-dependent) effective quantum numbers $\nu^*$ through $-\frac{1}{2\nu^{*2}}=-\frac{1}{2\nu^{2}}+\frac{1}{R}$ instead produces a small difference in the range $[0.02, 0.35]$ MHz for $r_{\rm LR}\in[50,150]$ a.u. in the perturbative long-range corrections of the $20p\Sigma$ state, which we neglect(this difference is even smaller for higher $n$). 

\begin{figure}[t]
\centering
\includegraphics[width=0.49\textwidth]{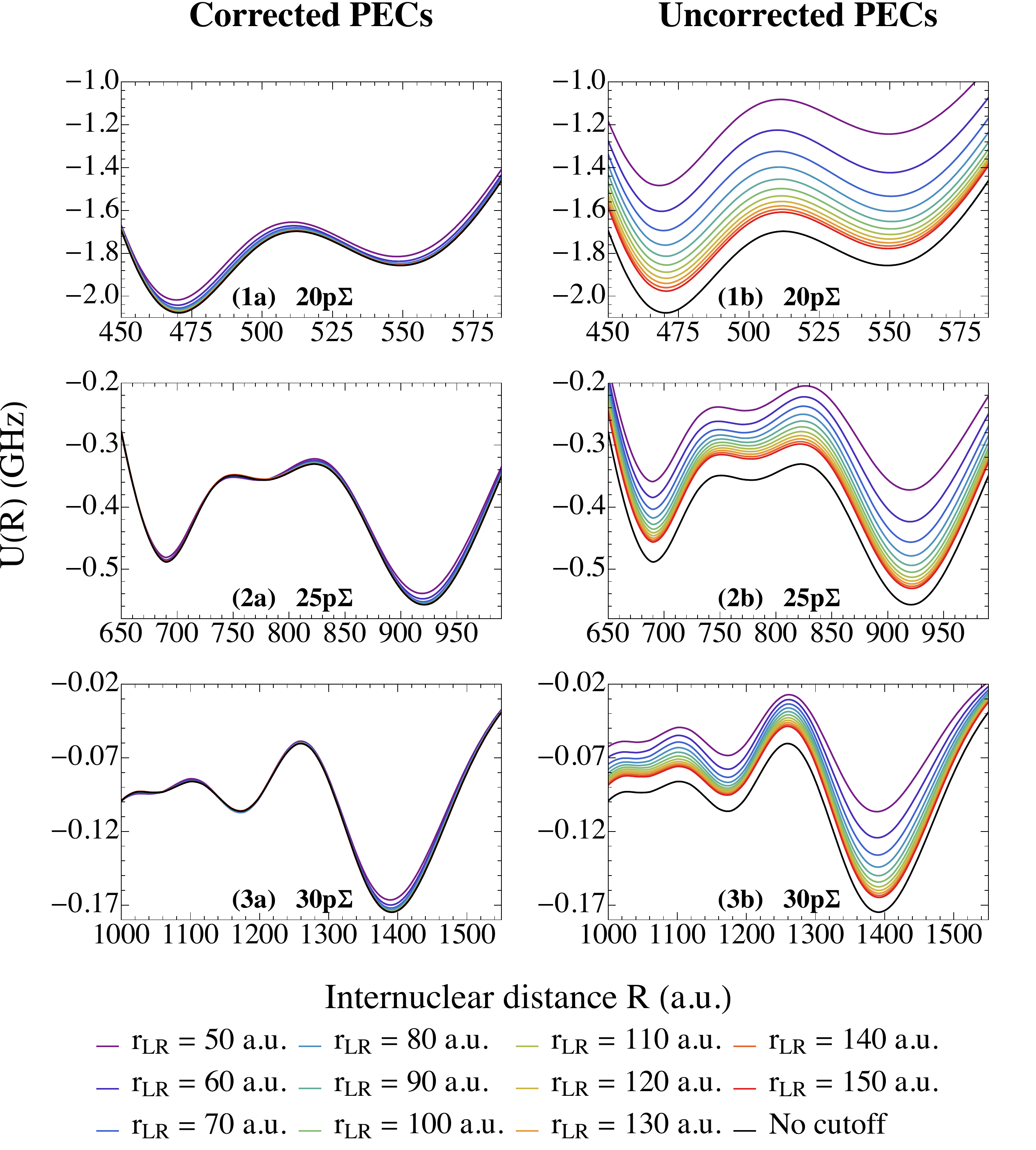}
\caption{
The panels (1,2,3) show the PECs near the outermost well of Rb($np$)-Rb($5s$) with molecular $\Sigma$ symmetry that asymptotically approach the $(20p,25p,30p)$ states, respectively. They are separated into two groups: right panels (b) that show the PECs of different cutoff radii that are calculated using full diagonalization, {\it i.e.} $\bar{U}_{\rm B}(R;r_{\rm LR})$, {\it without} the long-range perturbative correction and the left panels (a) that show the same PECs with the addition of the first order perturbative long-range corrections (calculated with Whittaker functions), $\bar{U}_{\rm B}(R;r_{\rm LR})+\expval{\Delta V_{\rm Sc}}(R;r_{\rm LR})$. The black curves, $U_B(R)$, that appears in both panels (a) and (b) are the PECs calculated with the no cutoff potential. All of the corrected PECs converge to the no cutoff PECs, namely $U_B(R)\approx \bar{U}_B(R;r_{\rm LR})+\expval{\Delta V_{\rm Sc}}(R;r_{\rm LR})$. This convergence gets better as the cutoff is increased because the long-range corrections become more perturbative for longer radii cutoffs, {\it i.e.} a manifestation of the limits  $\bar{U}_B(R;r_{\rm LR}\to\infty)=U_B(R)$ and $\expval{\Delta V_{\rm Sc}}(R;r_{\rm LR}\to\infty)=0$ by construction.}
\label{fig:BspPECs}
\end{figure}

\begin{figure}[h!]
\centering
\includegraphics[width=0.48\textwidth]{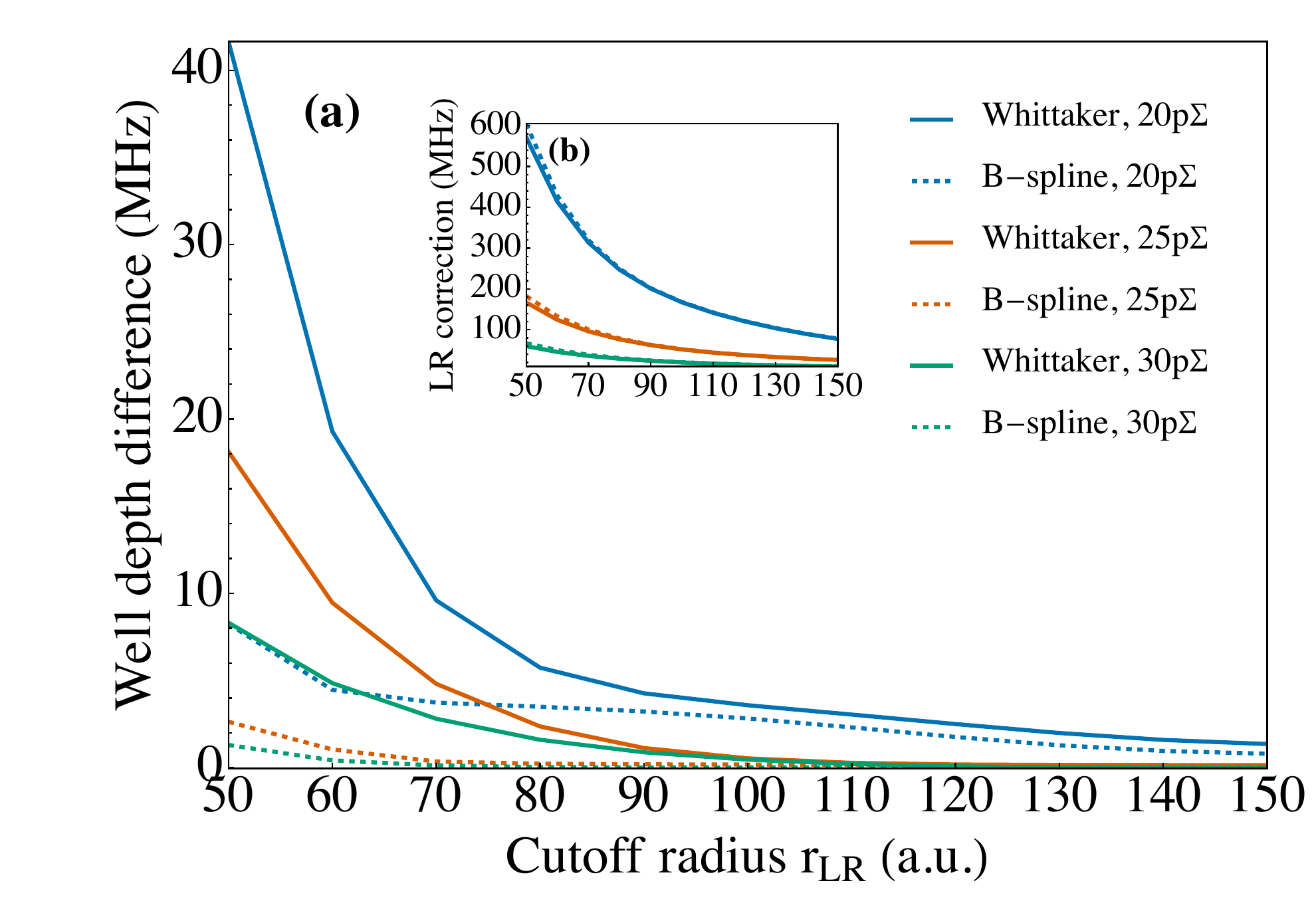}
\caption{The main figure (a) shows the magnitude of the outmost well(located at $R_{\rm out}$) depth difference, $\abs{U_{\rm B}(R_{\rm out})-\bar{U}_{\rm B}(R_{\rm out};r_{\rm LR})-\expval{\Delta V_{\rm Sc}}(R_{\rm out};r_{\rm LR})}$, of $\Sigma$ states of Rb($n p$)+Rb($5s$) and the inset figure (b) shows the long-range correction, $\abs{\expval{\Delta V_{\rm Sc}}(R_{\rm out};r_{\rm LR})}$, both in MHz for varying cutoff radius with steps of $10\, a_0$. The outer region corrections $\expval{\Delta V_{\rm Sc}}(R_{\rm out};r_{\rm LR})$ are calculated using the electronic wavefunctions of $n p\Sigma$ represented with Whittaker functions evaluated at unperturbed $e^-$-Rb($np$) energy levels and with the computed B-spline zeroth-order eigenfunction for the cutoff potential. Any well depth difference below 0.5 MHz is set to zero as it is beyond the accuracy of our full diagonalization computation. A good agreement in this figure does not suggest a valid cutoff immediately. One needs to use this figure and Fig.~\ref{fig:GFWells} together to find the optimal cutoff radius.}
\label{fig:BspConvLR}
\end{figure}

Fig.~\ref{fig:BspPECs} shows the outmost well of the PECs of $(20,25,30)p\Sigma$ as the lowest, middle, and highest in our analysis before and after the long-range corrections[calculated with eigenvectors in Eq.~\eqref{NormWhittaker}] are added. For the PECs in this figure, as well as the other ones for intermediate $n p\Sigma$, we can see that the perturbative correction for only sufficiently large cutoffs makes the PECs converge to the no cutoff PEC over the whole range of $R$, as shown in the same figure. Shorter $r_{\rm LR}$, such as 50 and 60 a.u. are presented to emphasize that the convergence is not guaranteed for all cutoff radii and may not reach the desired precision if it is chosen too short. The corrections calculated with the eigenvectors of the full diagonalization method are expected to produce more accurate results than Whittaker functions(but computationally much more costly) due to the self consistency as both $U_B(R),\bar{U}_B(R,r_{\rm LR})$ are calculated with full diagonalization. Corrections with full diagonalization eigenvectors are calculated for only the outmost well minima and are shown in Fig.~\ref{fig:BspConvLR} along with corrections calculated with Whittaker functions and the well depth difference between $U_B(R)$ and $\bar{U}_B(R;r_{\rm LR})+\expval{\Delta V_{\rm Sc}}(R;r_{\rm LR})$. This figure displays that the error in the perturbative corrections with Whittaker functions decreases significantly depending on the cutoff radius and reaches sub MHz levels for all $n$ around $r_{\rm LR}=90$. 

Moreover, Fig.~\ref{fig:BspConvLR} summarizes the full diagonalization results and presents a test of our overall strategy of replacing the solution of the full electronic Hamiltonian with a long-range polarization potential by instead solving the Schr\"odinger equation with the long-range potential cut off at some distance $r_{LR}$ from the perturber, followed by a perturbative treatment of the outer region that was initially cut off. It shows the long-range corrections $\expval{\Delta V_{\rm Sc}}(R;r_{\rm LR})$ calculated with the methods mentioned in the last paragraph and the depth difference between the outmost wells in Fig.~\ref{fig:BspPECs}, located at $R_{\rm out}$, {\it i.e.} $\abs{U_B(R_{\rm out})-\bar{U}_B(R_{\rm out};r_{\rm LR})-\expval{\Delta V_{\rm Sc}}(R;r_{\rm LR})}$. When either the cutoff radius or the energy is increased, the computed fixed-$R$ energy levels exhibit the expected convergence towards their values in the no cutoff case.

Naturally, if perturbation theory is used to incorporate the outer region corrections, the cutoff radius in the $e^-$-Rb($5s$) interaction must be chosen large enough such that the long-range corrections remain perturbative. However, an agreement between $U_{\rm B}(R)$ and $\bar{U}_{\rm B}(R;r_{\rm LR})+\expval{\Delta V_{\rm Sc}}(R;r_{\rm LR})$ that automatically occurs for long $r_{\rm LR}$, does not simply suggest picking an arbitrarily large cutoff radius, since it will have the best agreement in the diagonalization plus perturbation theory calculation. In fact, as we discuss in the next section, the size of $r_{\rm LR}$ determines the extent to which the zero-range pseudopotential based on phaseshifts in the cutoff potential can correctly replace the exact potential; thus, picking a cutoff radius too large makes phaseshift-based approaches such as the Fermi-Omont pseudopotential or the Green's function treatments of {\cite{eiles2023, SchmelcherGF} less accurate. On the other hand, if it is chosen too small, the contribution of the outer region that we take into account with perturbation theory becomes too large. If the problem limits one to pick short $r_{\rm LR}$ only where the $\expval{\Delta V_{\rm Sc}}$ is not perturbative anymore, one can employ a nonperturbative treatment like the outer region $R$-matrix theory~\cite{KenTaylorRef,grangerthesis,GrangerPRA} to account for the long-range correction.

\subsection{Potential energy curves of full diagonalization and the Green's function method}\label{chap:comments on on GFM}

In this section, we analyze the crucial steps (i), (ii) and elaborate on step (iiia) of our strategy by employing Fermi-Omont pseudopotential concept through a Green's function  method. There are different Green's function based treatments for calculating the PECs of Rydberg molecules~\cite{pruneleGF,khuskivadze2002adiabatic,hamilton2003photoionization,SchmelcherGF,eiles2023}. The present study incorporates the method developed in Ref.~\cite{hamilton2003photoionization} that utilizes the idea of the Fermi-Omont pseudopotential, which neglects the spin degrees of freedom. Somewhat later, in Ref.~\cite{eiles2023}, that method was generalized to incorporate electronic and nuclear spin interactions, but our present model study does not include that physics. In contrast to diagonalization methods, however, which use the Fermi-Omont electron-perturber interaction method as a term in the Hamiltonian, this Green's function method uses the bare $S$-wave and $P$-wave scattering phaseshifts to define boundary information on a tiny spherical surface centered on the perturber, giving a Kirchoff integral expression for the electronic eigenfunctions. An immediate implication of this assumption is that the scattering event is isolated from the core's reach. An advantage of the Green's function treatments is that they do not suffer from the erratic convergence properties that have been observed in applications of the diagonalization method. 

GFM, as an implementation of the Fermi-Omont pseudopotential concept, gives us a theoretical framework to investigate the exact potential's representability by the Fermi-Omont pseudopotential and the unphysical zero-energy divergence caused by using it in conjunction with long-range potential phase shifts that deviate from the Wigner threshold law. To implement our strategy, it is necessary to make a small modification to the transcendental equation of GFM to accept the energy-analytic phase shifts instead of the energy-normalized ones; this step is shown in Appendix~\ref{append:GFreform}.

\begin{figure*}[t]
\centering
\includegraphics[width=0.8\textwidth]{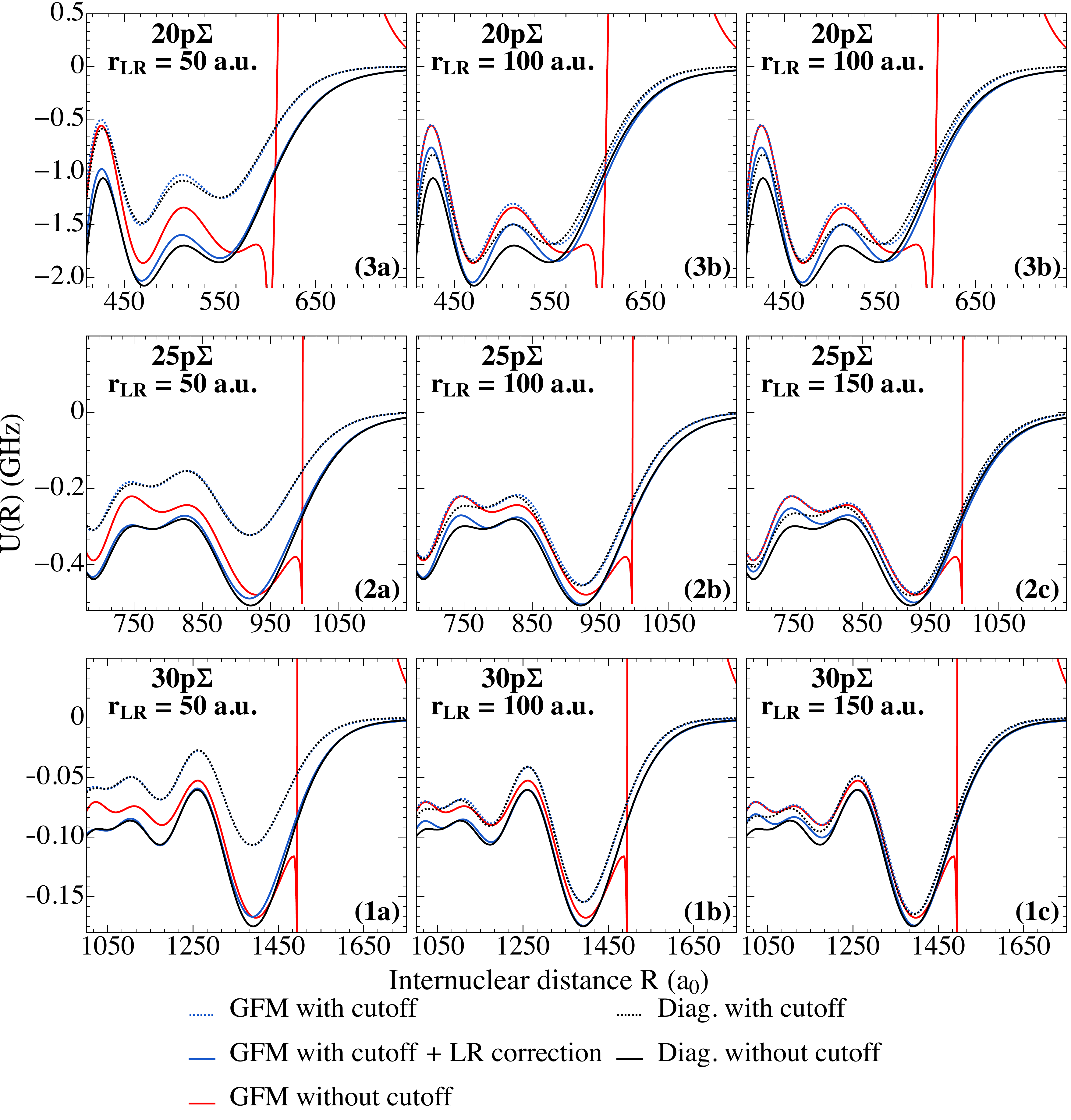}
\caption{Rows (1,2,3) show the PECs of asymptotically $(30p\Sigma,25p\Sigma,20p\Sigma)$ states, and columns (a), (b) and (c) have varying cutoff radii, $r_{\rm LR}=50,100,150$ a.u., respectively. The dashed curves, $U_B(R),\bar{U}_B(R;r_{\rm LR})$, are the results of the full diagonalization and the solid curves, $U_G(R),\bar{U}_G(R;r_{\rm LR}),\bar{U}_G(R;r_{\rm LR})+\expval{\Delta V_{\rm Sc}}(R;r_{\rm LR})$, are the GFM results. It is observed that, $U_{\rm B}(R)\approx\bar{U}_{\rm G}(R;r_{\rm LR}) + \expval{\Delta V_{\rm Sc}}(R;r_{\rm LR})$ is satisfied most accurately for $r_{\rm LR}=100$ a.u. in the $25p$ and $30p$ states. The optimum cutoff radius for the $20p$ state, $r_{\rm LR}=70$ a.u., results are presented in Fig.~\ref{fig:PECN20opt}. The disagreements in shorter internuclear distances are mostly due to $L\geq 2$ partial waves and are addressed in Fig.~\ref{fig:PECcorrN202530}. The disagreements in the outmost well depths in panels (a) are a consequence of taking the cutoff too short such that the long-range corrections become nonperturbative. These panel naturally exhibit the best agreement between $\bar{U}_B(R;r_{\rm LR})$ and $\bar{U}_G(R;r_{\rm LR})$ so it helps showing the transition of $\bar{U}_G(R;r_{\rm LR})$ from the diagonalization results to GFM results as the cutoff radius increases.}
\label{fig:PECcompN202530}
\end{figure*}

Fig.~\ref{fig:PECcompN202530} displays the PECs calculated with the GFM and full diagonalization using either the full-range or the cut off potential.
In the previous section, we showed that a short cutoff radius leads to a nonperturbative long-range correction $\expval{\Delta V_{\rm Sc}}(R;r_{\rm LR})$ and gives $U_B(R)\not\approx \bar{U}_B(R;r_{LR})+ \expval{\Delta V_{\rm Sc}}(R;r_{\rm LR})$. Consequently, $U_B(R)\not\approx \bar{U}_G(R;r_{LR})+ \expval{\Delta V_{\rm Sc}}(R;r_{\rm LR})$ because the pseudopotential is more accurate in replacing shorter range potentials; therefore, it produces compatible PECs, $ \bar{U}_B(R;r_{LR})\approx \bar{U}_G(R;r_{LR})$. Instead, as seen in Fig.~\ref{fig:PECcompN202530}, we show that a long cutoff radius can also lead to a disagreement $U_B(R)\not\approx \bar{U}_G(R;r_{LR})+ \expval{\Delta V_{\rm Sc}}(R;r_{\rm LR})$, but now the long-range corrections are accurate, so the disagreement is due to the inaccuracy of the pseudopotential. This is a good indication for the relation between the range of the exact potential and its suitability to be replaced by a pseudopotential, suggesting the existence of a range limit to the exact potential for this replacement to be accurate.

Then, there are two competing factors with their sizes depending on the range of the cutoff radius. As seen in the last section, if it is too short, the long-range corrections are nonperturbative while as seen in this section, if it is too long, the pseudopotential cannot replicate the exact potential accurately. Depending on whether $r_{\rm LR}$ is chosen too short or too long, one of these two competing factors starts to take over and leads to inaccuracies. As a consequence, an intermediate cutoff radius gives the best results. This behavior is displayed in Fig.~\ref{fig:PECcompN202530}, and it is especially apparent in the PECs of the asymptotically $20p\Sigma$ state shown in row (3).  Note that these figures do not show the optimum cutoff radius for $20p\Sigma$.

As seen in Fig.~\ref{fig:PECcompN202530}, $\bar{U}_G(R;r_{\rm LR})$ approaches $\bar{U}_B(R;r_{\rm LR})$ when the cutoff is sufficiently short. On the other hand, if the cutoff is set too long, then naturally $\bar{U}_G(R;r_{\rm LR})$ approaches $U_G(R)$ that suffer from replacing such long-range scattering potential and the divergence caused by it. This is a sign of a successful representation of the exact potential by the pseudopotential when its range is sufficiently short. Fig.~\ref{fig:GFWells} further presents the validity of the pseudopotential to replace the exact potential, for the cases shown in detail for the $(20,25,30)p\Sigma$ potential curves in Fig.~\ref{fig:PECcompN202530}. It shows the outmost well depth differences of $(20,25,30)p\Sigma$ calculated with GFM and full diagonalization. Both figures show that the representation of longer range potentials through a zero-range pseudopotential is worse, as expected, because it cannot accurately incorporate the physics in the long-range potential.

\begin{figure}[h!]
\centering
\includegraphics[width=0.45\textwidth]{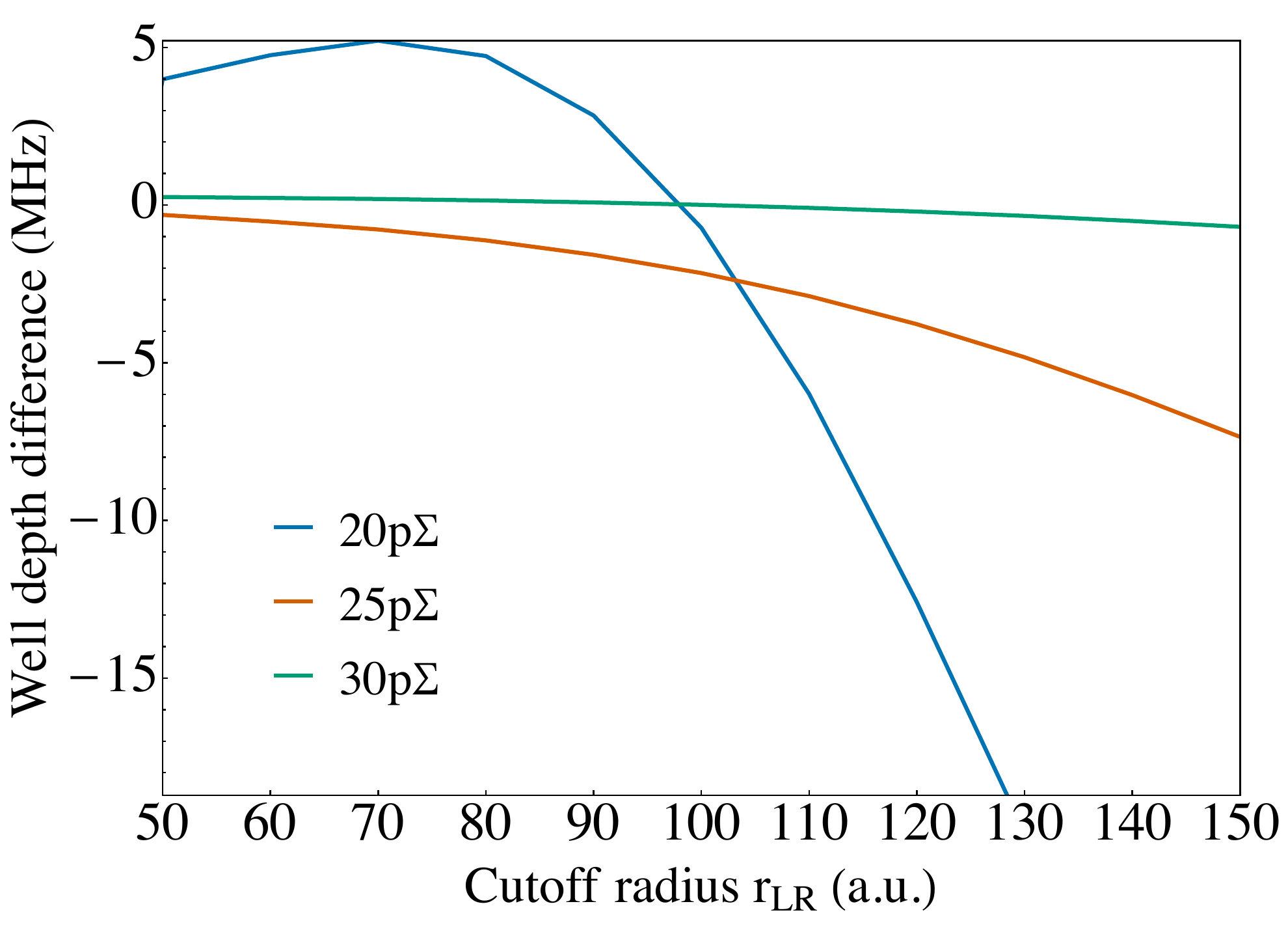}
\caption{The magnitude of the outmost well depth difference (located at $R_{\rm out}$), between the calculations using the GFM and the full diagonalization, $\bar{U}_{\rm B}(R_{\rm out};r_{\rm LR})-\bar{U}_{\rm G}(R_{\rm out};r_{\rm LR})$ of Rb($n p \Sigma$)+Rb($5s$) in MHz vs. cut radii ranging from $50 \, a_0$ to $150 \,a_0$ with steps of $10\,a_0$. Any difference below 0.5 MHz has been set to zero, since that is the limiting accuracy of our full diagonalization computation. Part of the disagreement between GFM and the full diagonalization results stems from the fact that $L\geq2$ partial waves in the $e^-$+Rb($5s$) scattering are neglected in the GFM treatment; their contributions are addressed in Fig.~\ref{fig:PECcorrN202530}. Again, a good agreement in this figure, observed in small cutoff radii, does not suggest a valid cutoff radius automatically because the long-range corrections for such short cutoff radius may be inaccurate as seen in Fig.~\ref{fig:BspPECs}. Furthermore, a short cutoff radius suppresses the disagreement that stems from the $D$- and $F$-waves corrections in the inner region as seen in Fig.~\ref{fig:PECcorrN202530}. Using this figure with Fig.~\ref{fig:BspConvLR} together is necessary to find the optimal cutoff.}
\label{fig:GFWells}
\end{figure}

\newpage 

We have shown the general picture of the new PECs after overcoming the polarizability pathology of the pseudopotential and obtained harmonious results between the two methods before and after the addition of the cut off long-range corrections in Fig.~\ref{fig:PECcompN202530}. In the next section, we apply the remaining step (iiib), {\it i.e.} add $D$- and $F$-wave corrections coming from the inner region($r<r_{\rm LR}$) that were neglected in this section. Then, we comment on the surviving disagreements that exist due to some of the assumptions made in GFM that have their roots in the zero-range interaction model.

\subsection{Comments on the remaining disagreement after correcting GFM}

The final PECs we obtained after applying our recipe are presented in Fig.~\ref{fig:PECcorrN202530}, along with the PECs obtained with the full diagonalization method using the no cutoff potential to show the agreement between the two different treatments of the same problem. The PECs before and after applying step (iiib) of our matching recipe are also presented in the same figure to display how the corrections coming from the $D$- and $F$-waves influence different regions of the PECs, based on the state $np$, the cutoff radius $r_{\rm LR}$ and the internuclear distance $R$.

\begin{figure*}[t]
\centering
\includegraphics[width=0.8\textwidth]{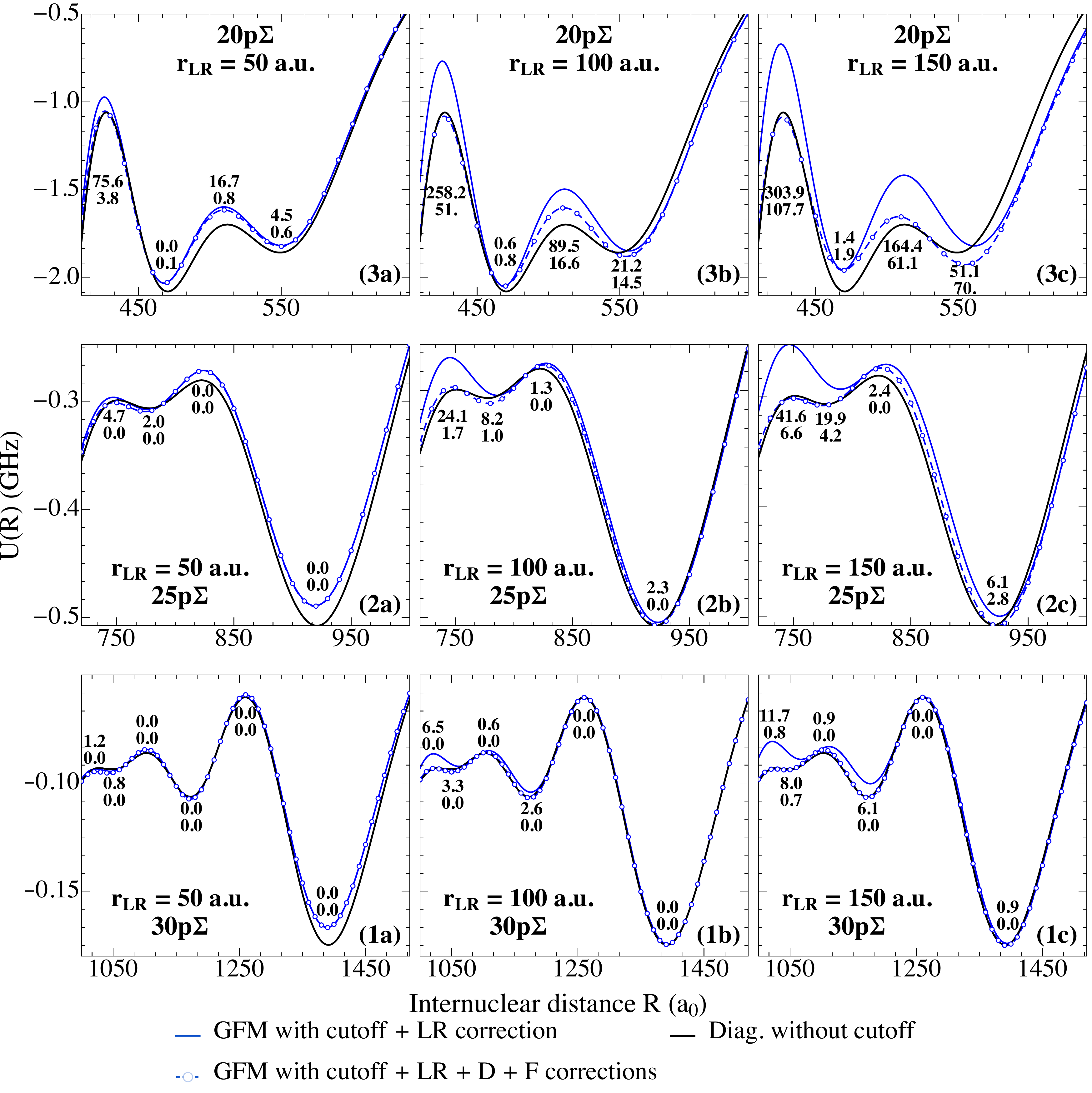}
\caption{This figure shows the PECs of asymptotically $20p,25p,30p$ states with molecular $\Sigma$ symmetry. $\bar{U}_G(R;r_{\rm LR})+\expval{\Delta V_{\rm Sc}}(R;r_{\rm LR})$(blue solid curve) accepts the perturbative $D$- and $F$-wave corrections to generate the blue dashed curve(with white points on it). Panels (a),(b) and (c) have $r_{\rm LR}=50,100,150$ a.u., respectively. The numbers at the points of peaks and dips are the $D$(top number)- and $F$(bottom number)-wave corrections at these points in MHz. The previously existing disagreement in shorter internuclear distances are partly resolved by the perturbative addition of the $D$- and $F$-wave corrections. The disagreements in the outmost well depth in panels (a) follow from the nonperturbative long-range corrections and not $L\geq 2$ waves, same as in Fig.~\ref{fig:BspPECs}. $20p\Sigma$ sees the $L\geq 2$ partial waves the strongest because electron's collision energy is largest in this state. If the cutoff is chosen sufficiently short, the addition of higher partial waves in the inner region($r<r_{\rm LR}$) makes a small correction, especially in the outmost well. The size of this correction depends on the electron's energy during the scattering, thereby it is less for states with a higher $n$ and at larger $R$. Panels (3) that show the asymptotically $20p\Sigma$ state displays worse agreement for large cutoff radii, $r_{\rm LR} = 100,150\, a_0$, which otherwise produces a good agreement for higher $n$.}
\label{fig:PECcorrN202530}
\end{figure*}

Fig.~\ref{fig:PECcorrN202530} does not show the best agreement between the two methods for $20p\Sigma$, but rather tracks the progress of the previously presented PECs in Figs.~\ref{fig:BspPECs} and \ref{fig:PECcompN202530} after adding $D$ and $F$-wave corrections. The optimum cutoff radii are presented in Table~\ref{tab:optimal_cutoff}. Since $r_{\rm LR}=100a_0$ is already the optimum one for the $25p\Sigma$ and $30p\Sigma$ states and was plotted in Fig.~\ref{fig:PECcorrN202530}, we present in Fig.~\ref{fig:PECN20opt} the PECs of the $20p\Sigma$ state with the optimum cutoff radius $r_{\rm LR}=70a_0$.

\begin{table}[h!]
\centering
\small
\renewcommand{\arraystretch}{1.}
\begin{tabular}{@{}c@{\hspace{1mm}}|@{\hspace{1mm}}c@{\hspace{1mm}}|@{\hspace{1mm}}c@{}}
\toprule
\shortstack{Asymptotic \\state, $np$} &
\shortstack{Optimum cutoff \\radius, $r_{\rm LR}$ (a.u.)} &
\shortstack{Volume of influence \\ratio,
$\left(\frac{r_{\rm LR}}{2\nu^2}\right)^3 \times 10^4$} \\
\midrule
$20p$ & 70  & 16 \\
$21p$ & 80  & 17 \\
$22p$ & 80  & 12 \\
$23p$ & 90  & 13 \\
$24p$ & 90  & 10 \\
$25p$ & 100 & 10 \\
$26p$ & 100 & 8  \\
$27p$ & 100 & 6  \\
$28p$ & 100 & 5  \\
$29p$ & 100 & 4  \\
$30p$ & 100 & 3  \\
\bottomrule
\end{tabular}
\caption{Optimal long-range cutoff radii for different
$np$ states and the corresponding volume of influence ratio.}
\label{tab:optimal_cutoff}
\end{table}

When GFM imposes the scattering information as a boundary condition on the electron at the perturber's position, it assumes a significantly smaller sphere of influence for the perturber, ${\rm Vol}_P$, relative to the core's, ${\rm Vol}_C$. Furthermore, it assumes that the scattering event is isolated from the core such that the Coulombic electron-core interaction is locally constant over the very small sphere of influence of the perturber. An estimate of the outmost well location is the previously divergent $R=2\nu^2$ point where the outmost well resides nearby; so, the ratio of their spheres of influence is $\frac{{\rm Vol}_P}{{\rm Vol}_C}\approx\left(\frac{r_{\rm LR}}{2\nu^2}\right)^3$ and they are given for optimum $r_{\rm LR}$ value of each $np$ state in Table~\ref{tab:optimal_cutoff}. It can help us to understand this qualitative behavior in a more quantitative way. 
The fact that the same cutoff radius leads to a reasonable agreement between GFM and full diagonalization for $25p$ and $30p$ but not for $20p$ suggests that the disagreement is partly about how different length scales of the collective three body system are competing depending on $n$. The size of the perturber's sphere of influence compared to the size of the molecule is larger in smaller $n$ states with the same cutoff radius. Therefore, the zero-range interaction model starts to fail at modeling the finite spatial extent of the electron-perturber interaction. So, this ratio tells us how realistic the GFM treatment is to ignore the far reach of the perturber when it imposes a boundary condition on the electronic wavefunction at zero distance from the perturber. When it gets large, {\it i.e.} when the sphere of influence of the perturber is considerably large compared to the core's, the imposed boundary condition on the electron ignores the much longer spatial extent of the exact electron-perturber scattering potential. Furthermore, around $r_{\rm LR}\sim 100$ a.u. near where the matching is done in Eq.~\eqref{engAnPS}, the assumption of locally constant Coulomb interaction between the electron and the core starts to be inaccurate for lower $n$ states, {\it e.g.} by $\abs{\frac{1/r-1/R}{1/R}}\approx 15\%$ for $20p$ while it is $9.8\%,6.7\%$ for $25p,30p$, respectively. The disagreement between the two methods increases simultaneously with this error and with the ratio $\left(\frac{r_{\rm LR}}{2\nu^2}\right)^3$. For a cutoff of $r_{\rm LR}=100a_0$, this ratio is 0.0046,0.001,0.0003 for the $20p,25p,30p$ states, respectively. Meaning that the perturber's spatial reach relative to the core covers approximately 15 times more volume for $20p$ than $30p$. If the cutoff radius is chosen as the optimum value for $20p$ which is $70a_0$, then this ratio is 0.0016, corresponding to a value approximately 5.3 times the optimum one for $30p$ with $r_{\rm LR}=100 \, a_0$. In Fig.~\ref{fig:PECN20opt}, we see that even this cutoff radius does not lead to an agreement as good as it is achieved for $30p$. The remaining disagreement at the peak around $R\sim 510 a_0$ is a result of multiple factors and can be addressed in the following ways: i) relaxing the assumption of locally constant Coulombic interaction between the electron and the core near the scattering event, ii) treating the excluded $L\geq 2$ partial waves with a more sophisticated treatment than first order perturbation theory.

\begin{figure}[h!]
\centering
\includegraphics[width=0.48\textwidth]{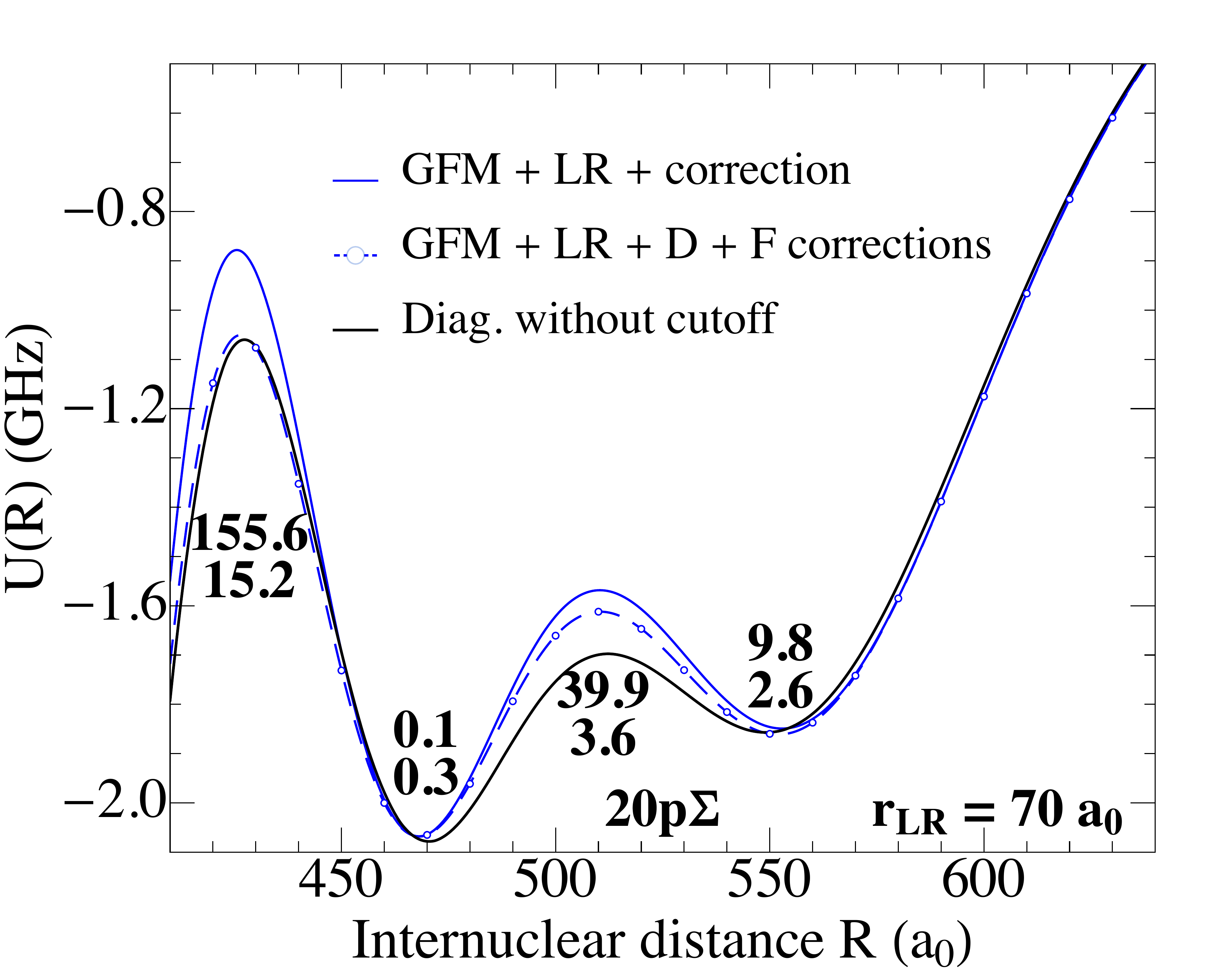}
\caption{This figure shows the PECs of asymptotically $20p\Sigma$ state, calculated with GFM(blue) and full diagonalization(black). The curves are the same with Fig.~\ref{fig:PECcorrN202530} except for $r_{\rm LR}=70a_0$. Out of the manifolds we have investigated in this paper, it has the worst overall agreement between GFM and full diagonalization. This is not surprising because it has the lowest $\nu$ in our analysis with the largest electron collision energy, hence $D$- and $F$-waves are felt stronger in this collision. Still a reasonable agreement is achieved in terms of the well depths and the first peak around $R\approx 425 a_0$ after adding the inner $D$- and $F$-wave corrections, but the difference in the second peak between the two outmost wells are not completely resolved. As shown in panels (3) of Fig.~\ref{fig:PECcorrN202530} and here, both of the peaks gets influenced by $L\geq 2$ partial wave scatterings which are neglected in the GFM treatment. The first peak naturally gets larger first order corrections as it is formed in smaller $R$ compared to the second one and yet it has a better agreement. Therefore, the disagreement in the second peak should be also due to other factors such as the assumption of locally constant Coulomb potential in GFM.}
\label{fig:PECN20opt}
\end{figure}

Furthermore, as seen in Table~\ref{tab:optimal_cutoff}, the optimum cutoff radius stops growing at one point. This makes sense and it suggests an optimum cutoff radius for $n\geq 25$, because the sphere of influence ratio goes as $\nu^{-2}$ and as it gets smaller the assumptions of GFM based on the Fermi-Omont pseudopotential idea discussed earlier in this section gets more realistic.

\newpage
\section{Conclusion}
Although zero-range pseudopotentials are very useful in simplification of complicated physical problems and therefore heavily utilized in many fields of physics, we emphasized in this paper that they have fundamental issues that one should consider; especially in the case of modeling long-range potentials and showed an Rb$_2$ model as an example case. We overcame the issue of representation of long-range potentials and the divergence caused by it by introducing a long-range cutoff function $f_{\rm LR}(r;r_{\rm LR})$ with the cutoff radius $r_{\rm LR}$, to the full-range scattering potential. As a benchmark test, we compared the results of the different treatments of the same scattering potential, one was a Green's function formalism that employs the pseudopotential concept and the other a full diagonalization with B-splines that uses the exact potential. Even though we used a certain Green's function method, the implications of our analysis here apply to methods that employ zero-range pseudopotentials modeling long-range potentials in general.

Each method provides two PECs with and without a cutoff, respectively: $\bar{U}_{\rm G}(R;r_{\rm LR}),U_{\rm G}(R)$ and $\bar{U}_{\rm B}(R;r_{\rm LR}),U_{\rm B}(R)$. As a consequence of the long-range representation issue, the PECs of the two different methods disagree when the potential is not cut at long-range, $U_{\rm G}(R)\not\approx U_{\rm B}(R)$, especially near zero electron collision energy, where $U_G(R)$ diverges at $R=2\nu^2$. We showed that introducing a cutoff to the infinite range polarizability component of the scattering potential removes this divergence. Moreover, by construction, we have $\bar{U}_{\rm G}(R,\infty)=U_{\rm G}(R)$. If the cutoff radius is short enough such that the long-range representation problem is resolved, we start to obtain $\bar{U}_{\rm G}(R;r_{\rm LR}) \approx \bar{U}_{\rm B}(R;r_{\rm LR})$. Next, we calculated the long-range corrections $\expval{\Delta V_{\rm Sc}}(R;r_{\rm LR})$ and verified that $U_{\rm B}(R) \approx \bar{U}_{\rm G}(R;r_{\rm LR}) + \expval{\Delta V_{\rm Sc}}(R;r_{\rm LR})$ for some $r_{\rm LR}$. We utilized perturbation theory in this paper as the Rb$_2$ system allows it in the energy regime we considered; but one can employ nonperturbative methods if necessary.

This strategy gives smooth PECs, $U_B(R)\approx U_G(R;r_{\rm LR})+\expval{\Delta V_{\rm sc}}(R;r_{\rm LR})$, eliminating the PEC cusps near zero electron collision energy that existed in some previous applications of both Green's function and diagonalization methods to ultralong-range Rydberg molecules. See, for instance, Fig.1 of each of Refs.\cite{Rost2010,sasmannshausen_experimental_2015}.

Furthermore, we show that there exists a range for the cutoff radii $r_{\rm LR}\in [70,100] $ a.u. where $U_{\rm B}(R) \approx \bar{U}_{\rm G}(R;r_{\rm LR}) + \expval{\Delta V_{\rm Sc}}(R;r_{\rm LR})$ is satisfied and the two methods agree. In principle, a cutoff radius that works for a low $n$ PEC should also work for a higher one because at the outmost well, electron kinetic energy gets smaller as $n$ increases so the scattering is more localized near the perturber, giving a more suitable system to be modeled by a pseudopotential. In other words, the optimum value of $r_{\rm LR}$ that yields the best agreement between the PECs of the two methods increases with an increase in $n$ until it saturates around $r_{\rm LR}=100\, a_0$ for $25p$ and beyond. This is because with the long-range cutoff function, as $n$ increases, the scattering event happens further away from the core in a fixed volume determined by $r_{\rm LR}$. Consequently, the assumptions that rely on how strictly the electron-core and electron-perturber are isolated from each other get strengthened.
This is also seen in the ratio of volumes of the spheres of influence between the perturber and the core, $\left(\frac{r_{\rm LR}}{2\nu^2}\right)^3$, getting smaller for higher $\nu$. Another consequence of having smaller kinetic energy at higher $n$ is having less discrepancy between $\bar{U}_G$ and $\bar{U}_B$ due to $L\geq 2$ partial waves which helps higher $n$ PECs to have a better agreement. Along the way, we estimated perturbatively the $D$- and $F$-wave pseudopotential contributions near the outmost well and show examples where they become considerable, especially in relatively low $n \sim 20$[see panels (3a,3b,3c) of Fig.~\ref{fig:PECcorrN202530}]. If one wants to make a precise calculation concerning the same and lower $n$ PECs than we investigated in this paper, $D$- and even $F$-waves should be taken into account, depending on the aimed precision.

Finally, after applying our strategy and considering the perturbative corrections, we obtained an overall good agreement between the GFM and full diagonalization, especially in relatively high $n$ PECs for $r_{\rm LR}\sim 100$ a.u. in the region where the two outmost wells reside. We tested it for lower $n$ states, down to $20p$, where we still get a good agreement for the two wells but not at the peak between them and discussed possible reasons for the remaining disagreement such as the disregard of $L\geq 2$ partial waves in GFM and the locally constant Coulombic $e^-$-Rb($np$) interaction assumption in the phase shift calculations. One can do a more quantitative analysis on the disagreement in this peak by taking into account the contributions of $L\geq 2$ waves as well as the contributions coming from their couplings. Another direction to investigate is the validity of the locally Coulombic electron-core potential assumption in the collision regime worked out in this paper by treating the electron-perturber scattering under the influence of the core through employing regular and irregular solutions in Eq.~\eqref{engAnPS} that see the variations in the Coulombic electron-core potential.

\textit{Acknowledgments: } 
This work has been supported in part by NSF Grant No. PHY-2512984.

\newpage

\appendix

\section{D- and F-wave contributions}\label{append:DF}

The contributions coming from $D$- and $F$-wave scatterings from the inner region($r<r_{\rm LR}$) are investigated with first order perturbation theory in this section. First, we invoke the $D$- and $F$-wave pseudopotentials from the terms of the expansion in Eq.~(2.18) derived in Ref.~\cite{omont1977theory},
\begin{equation}
    V_2(\vec{r})=-10\pi\delta^{{(3)}}(\vec{r}-R\hat{z})\frac{\tan\delta_2(R)}{k(R)}P_2\left(\frac{\overleftarrow{\nabla}\cdot\overrightarrow{\nabla}}{k^2(R)}\right),
\end{equation}
\begin{equation}
    V_3(\vec{r})=-14\pi\delta^{(3)}(\vec{r}-R\hat{z})\frac{\tan\delta_3(R)}{k(R)}P_3\left(\frac{\overleftarrow{\nabla}\cdot\overrightarrow{\nabla}}{k^2(R)}\right).
\end{equation}
where $P_n(\hat{x})$ are the Legendre polynomials. 
Before employing $V_{2,3}(\vec{r})$, following the derivation in the supplementary material of Ref.~\cite{Dpseudo}, we use the Schr\"odinger equation in the $R$ coordinate to manipulate them to the form that admits the scattering hypervolumes $a_2^5(\varepsilon), a_3^7(\varepsilon)$ associated with $D$- and $F$-wave scatterings.
\begin{equation}
    (V_2)_{i'i}=5\pi a_2^5(\varepsilon)[3(\nabla\cdot\nabla)^2-\nabla^2\nabla^2]\psi_{n'l'm'}(\vec{R}')\psi_{nlm}(\vec{R}).
\end{equation}
where $\psi_{nlm}(R)=\frac{u_{nl}(R)}{R}Y_{lm}(\hat{R})$ with spherical harmonics $Y_{lm}(\hat{R})$ and the reduced radial wavefunction $u_{nl}(R)$.

For $m=m'=0$ and $\vec{R}=R\hat{z}$, this simplifies to 
\begin{align}\label{Dsimple}
    (V_2)_{i'i}(R)&=\frac{5a_2^5(\varepsilon)}{R^6}v_{nl}(R)v^*_{n'l'}(R)\nonumber\\&=-\frac{5r_0^5\tan\delta_2^{\rm an}(\varepsilon)}{R^6}v_{nl}(R)v^*_{n'l'}(R),
\end{align}
where $v_{nl}(R)=\sqrt{l+1/2}\{[3+3l(l+1)/2-2R+(R/n)^2]u_{nl}(R)-3Ru'_{nl}(R)\}$\cite{Dpseudo}.

We do the same trick for the $F$-wave to get
\begin{align}
    (V_3)_{i'i}=7\pi a_3^7(\varepsilon)&\nonumber[5(\nabla\cdot\nabla)^3-3(\nabla\cdot\nabla)\nabla^2\nabla^2]\\&\psi_{n'l'm'}(\vec{R})\psi_{nlm}(\vec{R}).
\end{align}
From this form, at $\vec{R}=R\hat{z}$, we obtain
\begin{align}\label{Fsimple}
    (V_3)_{i'i}(R)&=\frac{7a_3^7(\varepsilon)}{R^8}w_{nl}(R)w_{n'l'}^*(R)\nonumber\\&=-\frac{7r_0^7\tan\delta_3^{\rm an}(\varepsilon)}{R^8}w_{nl}(R)w_{n'l'}^*(R),
\end{align}
for the $m=m'=0$ symmetry. Here, $w_{nl}(R)=\sqrt{l+1/2}\{[15+25l(l+1)/2-14R+6(R/n)^2]u_{nl}(R)-[15+5l(l+1)/2-2R+(R/n)^2]Ru'_{nl}(R)\}$. In our perturbation calculations, the wave function is set to the normalized solution of the radial Schr\"odinger equation for a Coulombic potential with non-integer effective principal quantum numbers $\nu_l(n)=n-\mu_l(n)$ that accepts atomic quantum defects. After omitting the argument and the subindex as $\nu_l(n)\equiv \nu$ for simplicity, this wave function that we have used in the perturbative correction calculations of this work is the normalized Whittaker functions $ u_{nl}(R)=\mathcal{W}_{\nu,l+\frac{1}{2}}(R)$ of Eq.~\eqref{NormWhittaker}.

Fig.~\ref{fig:DFphsh} show the energy-analytic phase shifts of the $D$- and $F$-waves versus kinetic energy of the electron, in the $e^-$-Rb($5s$) scattering modeled with the potential in Eq.~\eqref{ScatPot};
calculated the same way, {\it i.e.} single channel generalized quantum defect theory, with those presented in Figs.~\ref{fig:Sphsh} and \ref{fig:Pphsh} for the $S$- and $P$-waves:
\begin{figure}[h!]
\centering
\includegraphics[width=0.48\textwidth]{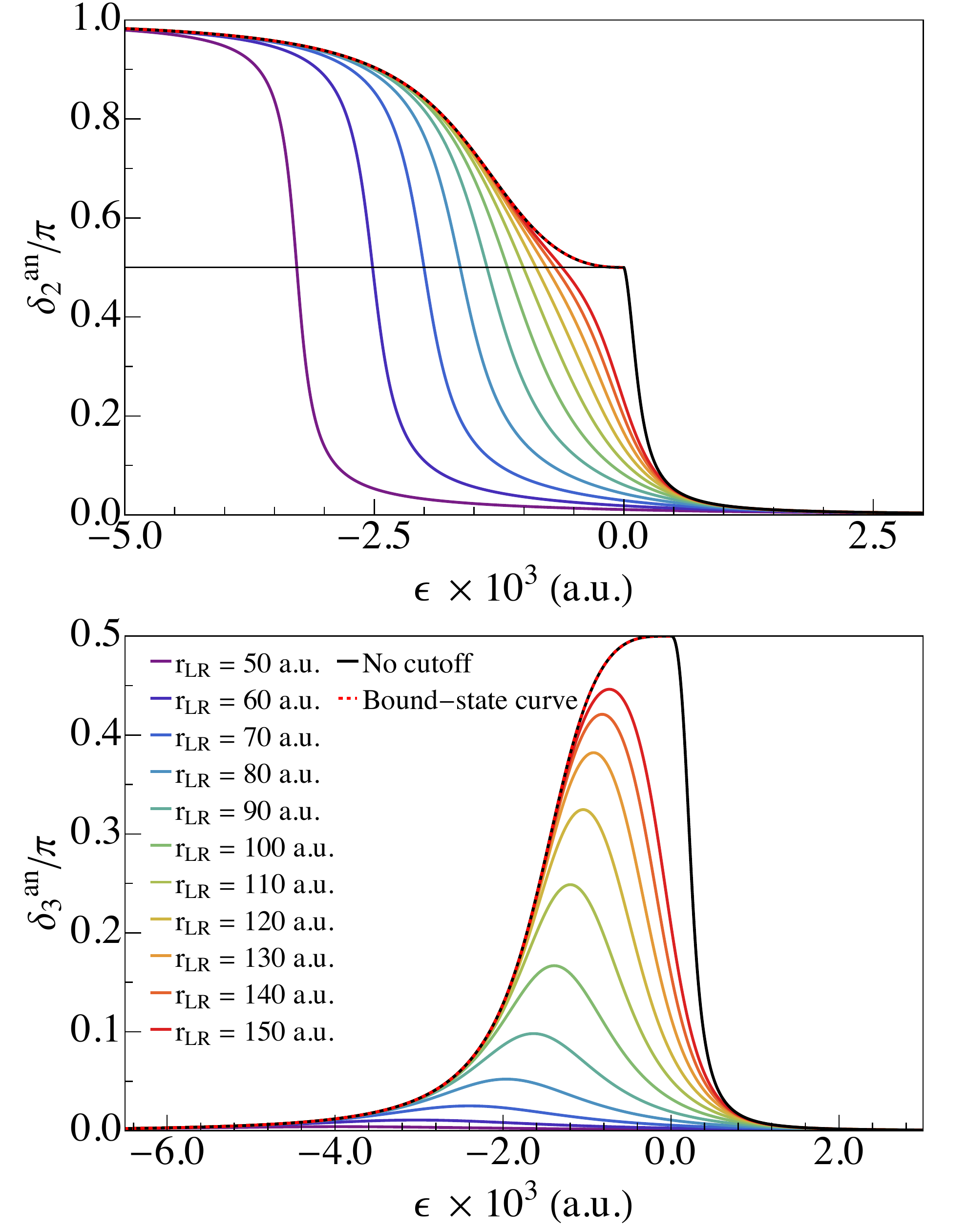}
\caption{The $D$ and $F$-wave energy-analytic phase shifts of $e^-$-Rb($5s$) subjected to the potential in Eq.~\eqref{ScatPot} with different long-range cutoff radii; calculated with Eq.~\eqref{engAnPS}.}
\label{fig:DFphsh}
\end{figure}

Again, the convergence to the no cutoff curve as the cutoff increases holds as expected.  The overall decrease in the phase shift based on the cutoff radii is more dramatic here than the $S$- and $P$- waves. As the cutoff decreases the energy range which these phase shifts become significant is pushed to very low energies that may seem irrelevant in realistic ULRM calculations in first glance. But looking at the no cutoff phase shift already tells us that the phase shift part that in Eqs.~\eqref{Dsimple} and \eqref{Fsimple} could be relevant. The $D$- and $F$-wave corrections to the PECs as written on the peaks/dips in Fig.~\ref{fig:PECcorrN202530} show that they already have a non-negligible influence on relatively low Rydberg levels of Rb$(n p)$-Rb$(5s)$ such as $20p$ and lose it as we go to higher $n$ values.

\section{Perturbative phase shift correction}

In this section, we derive the first order correction to the energy-normalized phase shift emerging from treating the long-range part $\Delta V_{\rm Sc}(r)$ of a mixed-range scattering potential $V_{\rm Sc}(r)$ perturbatively. Assume that the potential can be separated into its short- and long-range components, $V_{\rm Sc}(r)=\bar{V}_{\rm Sc}(r)+\Delta V_{\rm Sc}(r)$. In the Rb$(np)$+Rb$(5s)$ scattering, these potentials are simply $\bar{V}_{\rm Sc}=V_{\rm Sc}(r)f_{\rm LR}(r;r_{\rm LR})$ and $\Delta {V}_{\rm Sc}=V_{\rm Sc}(r)[1-f_{\rm LR}(r;r_{\rm LR})]$, namely the finite and infinite range components, respectively. 

Now consider the radial Schr\"odinger equation with unit mass in a.u.,
\begin{align}\label{Full schrodinger eq}
    \left(H_0(r)+\Delta V_{\rm Sc}(r)-\varepsilon\right)u(r)=0.
\end{align}
here
\begin{align}
    H_0(r)=-\frac{1}{2}\dv[2]{r}+\frac{L(L+1)}{2r^2}+\bar{V}_{\rm Sc}(r).
\end{align}
is the zeroth order Hamiltonian with only the short-range component. The zeroth-order solutions obey
\begin{align}
    \left(H_0(r)-\varepsilon\right)U^{(0)}(r)=0,
\end{align}
at short-range $r<\bar{r}$; where $\varepsilon=\frac{k^2}{2}$. And at long-range, beyond $r\geq \bar{r}$, where $\bar{V}_{\rm Sc}(r)$ is negligible, we simply recover the free Schr\"odinger equation. So, at long-range, the regular and irregular solutions can be expressed by a linear combination of the spherical Bessel functions: $f_L(k r)=\left(\frac{2}{\pi k}\right)^{\frac{1}{2}}krj_L(kr)$, $g_L(k r)=\left(\frac{2}{\pi k}\right)^{\frac{1}{2}}kry_L(kr)$. Note that these are energy-normalized solutions with the Wronskian, ${\rm Wr} (f_L,g_L)=\frac{2}{\pi}$. 

For $r > \bar{r}$, the zeroth-order long-range solutions can be recasted in their quantum defect theoryquantum defect theory forms, and the phase shift with respect to the free solutions as
\begin{align}
    F_L^{(0)}(k,r)\underset{r>\bar{r}}{\equiv}f_L(kr)\cos\delta^{(0)}_L-g_L(kr)\sin\delta^{(0)}_L,
\end{align}
\begin{align}
    G_L^{(0)}(k,r)\underset{r>\bar{r}}{\equiv} f_L(kr)\sin\delta^{(0)}_L+g_L(kr)\cos\delta^{(0)}_L.
\end{align}
with the asymptotic forms
\begin{align}\label{F asymp}
    F_L^{(0)}(k,r)\xrightarrow[r>\bar{r}]{}\left(\frac{2}{\pi k}\right)^{\frac{1}{2}}\sin(kr-\frac{L\pi}{2}+\delta_L^{(0)}),
\end{align}
\begin{align}\label{G asymp}
    G_L^{(0)}(k,r)\xrightarrow[r>\bar{r}]{} -\left(\frac{2}{\pi k}\right)^{\frac{1}{2}}\cos(kr-\frac{L\pi}{2}+\delta_L^{(0)}).
\end{align}
We can use these two to construct the Green's function that solves the inhomogeneous equation,
\begin{align}\label{Green's function eq}
    \left(H_0(r)-\varepsilon\right)\mathcal{G}_{k L}(r,r')=-\delta(r-r'),
\end{align}
The Green's function that satisfies the (dis)continuity conditions and that obeys the scattering boundary condition is given by
\begin{align}\label{Greens function}
    \mathcal{G}_{\varepsilon L}(r,r')=\pi F_L^{(0)}(r_<)G_L^{(0)}(r_>),
\end{align}
where $r_<$($r_>$) is the smaller(larger) of $r,r'$. Now, the Green's function that solves Eq.~\eqref{Green's function eq} is used for obtaining the regular solution that solves Eq.~\eqref{Full schrodinger eq},
\begin{align}
    u(r)=F_L^{(0)}(r)+\int_{\bar{r}}^\infty \mathcal{G}_{\varepsilon L}(r,r')\Delta V_{\rm Sc}(r')u(r')dr'.
\end{align}
If we proceed recursively by setting the solution inside the integral as the zeroth-order solution of Eq.~\eqref{Full schrodinger eq}, $u(r')\approx F_L^{(0)}(r')$, we obtain the first-order solution as
\begin{align}
    u(r)=F_L^{(0)}(r)+\int_{\bar{r}}^\infty \mathcal{G}_{\varepsilon L}(r,r'\Delta V_{\rm Sc}(r')F_L^{(0)}(r')dr'.
\end{align}
By plugging in \eqref{Greens function} with the asymptotic forms \eqref{F asymp},\eqref{G asymp} and matching with the long-range quantum defect theoryquantum defect theory form that admits the first order correction
\begin{align}\label{first order asymp form}
    u(r)\to\left(\frac{2}{\pi k}\right)^{\frac{1}{2}}\Bigg[&\sin(kr-\frac{L\pi}{2}+\delta_L^{(0)})\nonumber\\&+\tan\delta_L^{(1)}\cos(kr-\frac{L\pi}{2}+\delta_L^{(0)})\Bigg].
\end{align}
at long-range, the first-order perturbative correction to the phase shift is found as
\begin{widetext}
\begin{align}\label{pert phaseshift formula}
    \tan\delta_L^{(1)}(
    k)&=-\pi\int_{\bar{r}}^\infty \left[f_L(kr)\cos\delta^{(0)}_L(k)-g_L(kr)\sin\delta^{(0)}_L(k)\right]^2 \Delta V_{\rm Sc}(r)dr\nonumber\\
    &=-2k\int_{\bar{r}}^\infty \left[j_L(kr)\cos\delta^{(0)}_L(k)-y_L(kr)\sin\delta^{(0)}_L(k)\right]^2 \Delta V_{\rm Sc}(r)r^2dr,
\end{align}
\end{widetext}
where we plugged in the regular and irregular free particle solutions in the last step. Fig.~\ref{fig:SPpertphsh} shows the harmony between the numerical calculation and the perturbative formula. In the numerical calculation, the lower integral limit $\bar{r}$ is set to a value slightly smaller than $r_{\rm LR}-\Delta r$ where $\Delta r=5$ $a_0$ is the width of the cutoff.  Notice that the short-range cutoff phaseshifts obey the expected Wigner threshold law $\delta_L^{(0)} \propto k^{2L+1}$ while the long-range perturbative corrections obey the polarization-modified threshold law at low energy, $\delta_L^{(1)} \propto k^{2-\delta_{L0}}$ where $\delta_{L0}$ is the Kronecker delta. These comparisons between the exact long-range phase shifts and the perturbative approximation are shown here to validate our recommended strategy discussed below of computing the long-range polarization corrections using perturbation theory. However, these long-range phaseshift corrections are not used in our final calculations of Rydberg molecule potential curves; rather the Rydberg electron wavefunction is used to compute that perturbation correction to the potential energy curve directly, instead of attempting to incorporate that long-range polarization physics in the Fermi-Omont pseudopotential.
\begin{figure}[h!]
\centering
\includegraphics[width=0.45\textwidth]{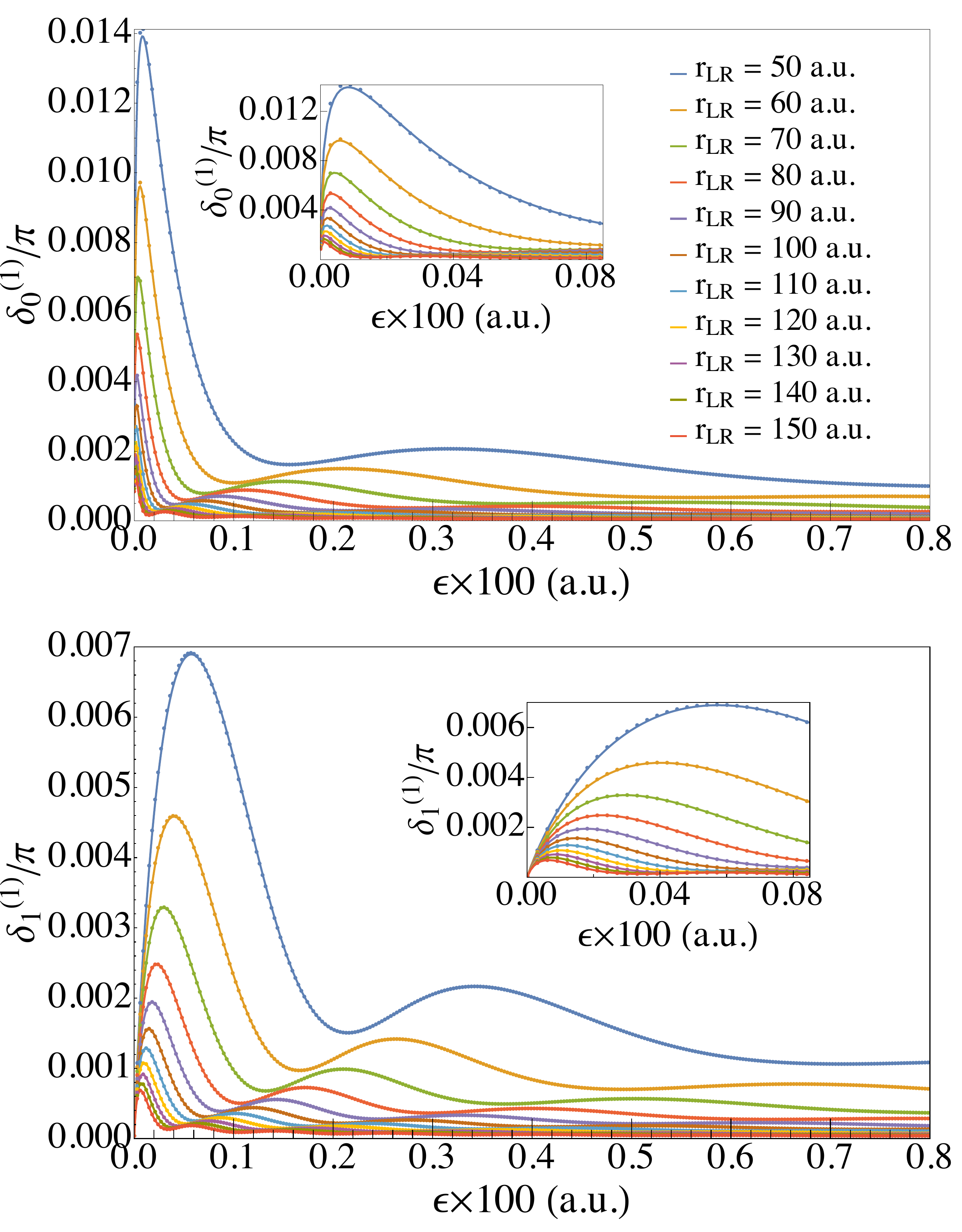}
\caption{The comparison of the perturbative formula (solid curves) and the numerical results (dots) for the $S$-(top panel) and $P$(bottom panel)-waves. The numerical results are calculated by taking the difference between the energy-normalized phase shifts of infinite and finite cutoff.}
\label{fig:SPpertphsh}
\end{figure}

The oscillations in Fig.~\ref{fig:SPpertphsh} are caused by the wave function term in the integrand of Eq.~\eqref{pert phaseshift formula}. The potential is independent of energy; thus, variations in energy are controlling what part of the wave function overlaps with the strongest part of the potential. For example, at the minima(maxima), the long-range potential overlaps with the wavefunction nodes(antinodes).

\section{Reformulation of GFM in terms of energy-analytic phase shifts}\label{append:GFreform}
The final set of linear equations of our GFM treatment, namely Eq. 32 of Ref.~\cite{eiles2023} can be readily modified to employ the energy-analytic phase shifts instead of the energy-normalized phase shifts. It is possible to achieve this by rescaling the determinantal equation in question by $\cos\delta^{L}_{\rm an}/\cos\delta^L$. In the spin dependent formalism of the reference, this factor would be $\cos\delta_{\rm an}^{S_\alpha,L_\alpha,J_\alpha}/\cos\delta^{S_\alpha,L_\alpha,J_\alpha}$. After rescaling, the energy-normalized phase shift factors that appear in the matrices in Eqs. (32) and (33) of Ref.~\cite{eiles2023} become $\cos\delta^{S_\alpha,L_\alpha,J_\alpha} \to \cos\delta_{\rm an}^{S_\alpha,L_\alpha,J_\alpha}$ and $\sin\delta^{S_\alpha,L_\alpha,J_\alpha}/k^{2L_\alpha+1}\to r_0^{2L_\alpha+1}\sin\delta_{\rm an}^{S_\alpha,L_\alpha,J_\alpha}$. With this transformation, the new determinantal equation is reformulated to admit the energy-analytic phase shifts while keeping its roots invariant. 

For simplicity, we present here the final transcendental equation after doing this manipulation. In Eq.~(D7) of Ref.~\cite{eiles2023}, the transcendental equation of molecular $\Sigma$ states for the spinless case is given as
\begin{align}\label{transEqBefore}
    0=&\left(1-\Phi_v \frac{\tan\delta_0}{k}\right)\bigg[1+(\Phi_{vvv}-3\Phi_{uuv})\frac{\tan\delta_1}{k^3}\nonumber\\&-\frac{\tan\delta_0/k}{1-\Phi_v\tan\delta_0/k} 3\Phi^2_{u v}\frac{\tan\delta_1}{k^3}\bigg],
\end{align}
which agrees with the derivations in Ref.~\cite{hamilton2003photoionization}. Here, the subscripts of $\Phi$ terms denote the partial derivative of the Green's function with respect to that variable. The details of the derivation of Eq.~\eqref{transEqBefore} are beyond the scope of this study and they can be found in Ref.~\cite{eiles2023}. All of the phase shifts that appear in Eq.~\eqref{transEqBefore} are energy-normalized, and they are defined only for positive energies as mentioned in the main text. They also suffer from the divergence at zero collision energy if the potential is long-range. Doing the aforementioned scaling replaces the energy-normalized phase shifts with energy-analytic counterparts so it gives
\begin{widetext}
\begin{equation}\label{transEqAfter}
    0=\left(\cos\delta_0^{\rm an}-\Phi_v r_0\sin\delta^{\rm an}_0\right)\bigg[\cos\delta_1^{\rm an}+(\Phi_{vvv}-3\Phi_{uuv})r_0^3\sin\delta^{\rm an}_1-\frac{r_0\sin\delta_0^{\rm an}}{\cos\delta_0^{\rm an}-\Phi_vr_0\sin\delta_0^{\rm an}} 3\Phi^2_{u v}r_0^3\sin\delta^{\rm an}_1\bigg].
\end{equation}
\end{widetext}
By performing this modification of the determinantal GFM equation, potential energy curves can be computed even at small negative electron-perturber collision energies. This is a crucial step for eliminating the potential curve cusps that show up at zero electron scattering energy that are observed in some pseudopotential based applications to ultralong-range Rydberg molecules.

\section{B-spline basis setup}\label{append:BspBasis}

Diagonalization with B-splines is carried out in polar coordinates centered on the nucleus of the ion, {\it i.e.} with only radial and polar angle coordinates since we consider here only azimuthally symmetric electronic $\Sigma$ states. There are (560,350) elements in the $(r,\theta)$ coordinates with degrees $8$ and $4$, respectively, corresponding to a total of (577,359) knots. The radial elements are distributed in the intervals $[R_i,R_f]$ as $r_i=R_i+(R_f-R_i)p_i^\alpha$, $p_i=i/N$, $i\in[1,N]$ where $N$ is the number of elements in this interval, with $\alpha=3$ near the core($r\in [0,20]$ a.u.) and the perturber($r\in [R-10,R+10]$ a.u.), and $\alpha=2$ in the intermediate region and at long distances.  The angular knots are distributed similarly, $\theta_j=\pi p_j^\alpha$ ,$j\in[1,N]$ with $\alpha=2$ in the interval $\theta\in[0,0.1]$ and linearly outside. We apply Dirichlet boundary condition on both ends of the radial coordinate and arbitrary boundary condition in the angular coordinate. As a result, this setup generates (566,354) B-spline basis elements in ($r,\theta$) coordinates. The knot distribution near the perturber is appropriately symmetrized about the perturber's position $\vec{R}=(R,0)$ by defining the knot sequence as $r_i=R_i+(R_f-R_i)[1-(1-p_i)^\alpha]$ in the near perturber region between the core and the perturber. In summary, these basis elements are placed denser near the core and the perturber in each direction compared to the regions far from the atomic cores to meet the required resolution.
\bibliography{refs}
\end{document}